\pdfoutput=1
\documentclass[cernpreprint, atlasdraft=false, UKenglish, texmf, orcidlogo]{atlasdoc}
\usepackage{atlaspackage}
\usepackage{atlasbiblatex}

\usepackage[snippets=true]{atlasphysics}
\graphicspath{{logos/}{figures/}}

\usepackage{ANA-SUSY-2023-26-PAPER-defs}
\usepackage{adjustbox}
\usepackage{multirow}
\usepackage{makecell}
\usepackage{duckuments}

\AtlasTitle{Search for supersymmetry using vector boson fusion signatures and missing transverse momentum in $pp$ collisions at $\sqrt{s}=13$ TeV with the ATLAS detector}

\AtlasAbstract{%
This paper presents a search for supersymmetric particles in models with highly compressed mass spectra, in events consistent with being produced through vector boson fusion.
The search uses 140~fb$^{-1}$ of proton--proton collision data at $\sqrt{s}=13$ TeV collected by the ATLAS experiment at the Large Hadron Collider.
Events containing at least two jets with a large gap in pseudorapidity, large missing transverse momentum, and no reconstructed leptons are selected.  A boosted decision tree is used to separate events consistent with the production of supersymmetric particles from those due to Standard Model backgrounds.
The data are found to be consistent with Standard Model predictions.
The results are interpreted using simplified models of $R$-parity-conserving supersymmetry in which the lightest supersymmetric partner is a bino-like neutralino with a mass similar to that of the lightest chargino and second-to-lightest neutralino, both of which are wino-like.
Lower limits at 95\% confidence level on the masses of next-to-lightest supersymmetric partners in this simplified model are established between $117$ and $120$ GeV when the lightest supersymmetric partners are within $1$ GeV in mass.
}

\AtlasRefCode{SUSY-2023-26}

\PreprintIdNumber{CERN-EP-2024-246}

\AtlasJournal{JHEP}
\AtlasJournalRef{JHEP 12 (2024) 116}
\AtlasDOI{10.1007/JHEP12(2024)116}

\hypersetup{pdftitle={ATLAS document},pdfauthor={The ATLAS Collaboration}}

\begin{document}

\maketitle

\tableofcontents

\section{Introduction}
Weak interactions at hadron colliders offer both challenges and opportunities in searches for physics beyond the Standard Model (SM).  Despite their small cross-sections and low-momentum final states when compared with strong interactions, the signatures of weak interactions can enable discoveries of rare processes, such as: single-boson production through vector boson fusion (VBF)~\cite{STDM-2016-09,STDM-2017-27,CMS-SMP-17-011}; diboson production through vector-boson scattering (VBS)~\cite{STDM-2017-23,STDM-2017-26,STDM-2017-19,CMS-SMP-20-016,CMS-SMP-20-013,CMS-SMP-21-001,CMS-SMP-21-011,STDM-2018-32,STDM-2018-31,STDM-2022-06}; and triboson production~\cite{STDM-2017-22,CMS-SMP-19-014,CMS-SMP-19-013,STDM-2019-09,STDM-2021-09,mySTDM-2019-17,STDM-2018-33,myCMS-SMP-22-006}.  Pure-electroweak processes, such as VBS and VBF, have emerged as powerful new tools, used most notably in studies of the Higgs boson (see e.g. Ref.~\cite{HIGG-2023-11}).  The extremely low production cross-sections of such processes, typically smaller than a picobarn for pair-production of weak-scale particles, are mitigated by final states with a unique signature of two high-momentum jets with large separation in rapidity.  The large amount of collision data available from the CERN Large Hadron Collider (LHC) increasingly enables the use of such weak production modes, and their distinctive features, in searches for physics beyond the Standard Model.

One extension of the Standard Model that has so far eluded discovery at the LHC is Supersymmetry (SUSY)~\cite{Golfand:1971iw,Volkov:1973ix,Wess:1974tw,Wess:1974jb,Ferrara:1974pu,Salam:1974ig}, which predicts new particles that have identical quantum numbers to their SM partners with the exception of spin, with SM fermions having bosonic partners and SM bosons having fermionic partners.   Problems such as those related to naturalness~\cite{Barbieri:1987fn,deCarlos:1993yy} and the nature of dark matter motivate extensions of the SM like SUSY that are intimately connected with electroweak symmetry breaking.

SUSY models that address the naturalness problem often have some SUSY partners with masses near the electroweak scale.  Models with weakly interacting massive particle (WIMP) dark matter candidates can avoid overabundance of dark matter through co-annihilation mechanisms, which arise in SUSY models with compressed mass spectra that feature SUSY partners with nearly degenerate masses~\cite{Griest:1990kh,Edsjo:1997bg,duan2018probing}. SUSY scenarios featuring compressed mass spectra can result in collisions at the LHC in which SUSY particles are produced, but their SM decay products have such low momenta that they avoid detection.  In $R$-parity conserving SUSY models~\cite{Farrar:1978xj} the lightest supersymmetric particles (LSPs) are stable and often carry no electric charge, presenting viable dark matter candidates~\cite{Goldberg:1983nd,myEllis:1983ew}.  Since they do not interact with the detector, such LSPs are only indirectly observed through the momentum imbalance in the transverse plane left by their escape.  Detection of a SUSY signal in such events is challenged by large backgrounds and, in models where the lightest SUSY partners are the SUSY counterparts of SM electroweak bosons, by low signal cross-sections.  One strategy for improving the sensitivity of searches for low-mass electroweak SUSY is to require additional hadronic activity in the event, which has the dual effect of providing a significant Lorentz boost to the SUSY particles and adding another visible physics object to the final state.  The resulting signature provides better separation of the SUSY signal from SM backgrounds.

\begin{figure}[!htb]
\centering
\subfloat[]{
\includegraphics[width=0.4\columnwidth]{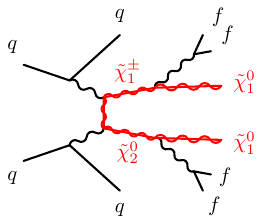}
\label{fig:feyn-VBF}
}
\hspace*{1cm}
\subfloat[]{
\includegraphics[width=0.4\columnwidth]{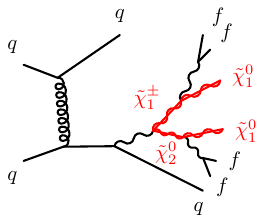}
\label{fig:feyn-DY}
}
\caption{Diagrams illustrating both (a) pure-electroweak and (b) mixed electroweak-QCD production of electroweakino pairs in association with two jets.}
\end{figure}

Following this strategy, this paper presents a search for compressed, electroweak SUSY in events with large missing transverse momentum, two or more jets rendering a VBF topology, and no reconstructed electrons or muons.  The search is optimised for SUSY models with LSP masses within 2~\GeV{} of the next-to-lightest SUSY partner (NLSP) mass, where the LSP is taken to be a bino-like neutralino (\chioz), and the mass-degenerate NLSPs include a neutralino (\chitz) and a pair of charginos (\chiopm), all of which are wino-like.  A broad collection of signature-based searches was studied in the context of the phenomenological minimal supersymmetric extension of the SM (pMSSM) in Ref.~\cite{SUSY-2020-15}, and the region with $\Delta m(\chitz,\chioz)<2~\GeV$ was found to be poorly constrained, motivating a focus on this region of SUSY parameter space. The \chitz, \chiopm and \chioz{} states are collectively referred to as electroweakinos in the following. The decays of unstable electroweakinos in this simplified model are assumed to proceed through emission of off-shell $W$ or $Z$ bosons, which then decay into fermions.  At such small mass splittings the final decay products of the \chitz{} and \chiopm{} have sufficiently low momenta that they are not reconstructed in the detector, motivating the zero-lepton signature.  The two jets can arise from either QCD couplings or pure-electroweak couplings.  In the pure-electroweak case, the SUSY particles can be produced through a VBF process, illustrated in Figure~\ref{fig:feyn-VBF}.  Other diagrams with QCD couplings, such as Figure~\ref{fig:feyn-DY}, result in the same final state and interfere with the VBF process.  The contributions of the VBF diagrams motivate searches in events characterised by large separation between jets and large dijet invariant mass.

The use of the VBF signature as a way to probe SUSY models has been proposed several times~\cite{Dutta:2012xe,Delannoy:2013ata,Dutta:2014jda,Cardona:2021ebw}.  This approach was explored experimentally by the CMS experiment in proton-proton ($pp$) collision data at $\sqrt{s}=8~\TeV$ in Ref.~\cite{CMS-SUS-14-005}, and at $\sqrt{s}=13~\TeV$ in Ref.~\cite{CMS-SUS-17-007}.  ATLAS studied the VBF final state in $pp$ collision data at $\sqrt{s}=13~\TeV$ in a two-lepton search for SUSY in Ref.~\cite{SUSY-2018-16}, and CMS presented SUSY searches in one- and zero-lepton topologies in Ref.~\cite{CMS-SUS-17-007}.  Similar final states have also been used to constrain the branching ratio of Higgs bosons to invisible final states~\cite{EXOT-2020-11,CMS-HIG-20-003}.
The kinematic selections applied in the latter analyses were optimised for the production of a single SM Higgs boson, rendering them less sensitive to the SUSY scenarios considered in this paper in which fermionic electroweakinos are pair-produced with an overall mass of roughly twice the mass of the SM Higgs boson. ATLAS has also performed a differential cross-section measurement of dijet kinematics in events with missing transverse momentum in VBF phase space~\cite{mySTDM-2018-55}, which observed good modelling of most kinematic distributions except the dijet invariant mass, partially motivating dedicated searches for non-SM contributions in this signature.

\section{ATLAS detector}

The ATLAS detector~\cite{PERF-2007-01} at the LHC covers nearly the entire solid angle around the collision point.\footnote{\AtlasCoordFootnote}
It consists of an inner tracking detector surrounded by a thin superconducting solenoid, electromagnetic and hadronic calorimeters,
and a muon spectrometer incorporating three large superconducting air-core toroidal magnets.

The inner-detector system (ID) is immersed in a \qty{2}{\tesla} axial magnetic field
and provides charged-particle tracking in the range \(|\eta| < 2.5\).
The high-granularity silicon pixel detector covers the vertex region and typically provides four measurements per track,
the first hit generally being in the insertable B-layer (IBL) installed before Run~2~\cite{ATLAS-TDR-19,PIX-2018-001}.
It is followed by the SemiConductor Tracker (SCT), which usually provides eight measurements per track.
These silicon detectors are complemented by the transition radiation tracker (TRT),
which enables radially extended track reconstruction up to \(|\eta| = 2.0\).
The TRT also provides electron identification information
based on the fraction of hits (typically 30 in total) above a higher energy-deposit threshold corresponding to transition radiation.

The calorimeter system covers the pseudorapidity range \(|\eta| < 4.9\).
Within the region \(|\eta|< 3.2\), electromagnetic calorimetry is provided by barrel and
endcap high-granularity lead/liquid-argon (LAr) calorimeters,
with an additional thin LAr presampler covering \(|\eta| < 1.8\)
to correct for energy loss in material upstream of the calorimeters.
Hadronic calorimetry is provided by the steel/scintillator-tile calorimeter,
segmented into three barrel structures within \(|\eta| < 1.7\), and two copper/LAr hadronic endcap calorimeters.
The solid angle coverage is completed with forward copper/LAr and tungsten/LAr calorimeter modules
optimised for electromagnetic and hadronic energy measurements respectively.

The muon spectrometer (MS) comprises separate trigger and
high-precision tracking chambers measuring the deflection of muons in a magnetic field generated by the superconducting air-core toroidal magnets.
The field integral of the toroids ranges between \num{2.0} and \qty{6.0}{\tesla\metre}
across most of the detector.
Three layers of precision chambers, each consisting of layers of monitored drift tubes, cover the region \(|\eta| < 2.7\),
complemented by cathode-strip chambers in the forward region, where the background is highest.
The muon trigger system covers the range \(|\eta| < 2.4\) with resistive-plate chambers in the barrel, and thin-gap chambers in the endcap regions.

The luminosity is measured mainly by the LUCID--2~\cite{LUCID2} detector that records Cherenkov light produced in the quartz windows of photomultipliers located close to the beampipe.

Events are selected by the first-level trigger system implemented in custom hardware,
followed by selections made by algorithms implemented in software in the high-level trigger~\cite{TRIG-2016-01}.
The first-level trigger accepts events from the \qty{40}{\MHz} bunch crossings at a rate below \qty{100}{\kHz},
which the high-level trigger further reduces in order to record complete events to disk at about \qty{1}{\kHz}.

A software suite~\cite{SOFT-2022-02} is used in data simulation, in the reconstruction
and analysis of real and simulated data, in detector operations, and in the trigger and data acquisition
systems of the experiment.

\section{Data and simulated event samples}
\label{sec:samples}

Data events were selected with a missing transverse momentum (\ptmiss) trigger~\cite{TRIG-2019-01}, employing varied trigger thresholds depending on the data-taking periods.
The trigger is $>95\%$ efficient in all data-taking periods for events where the offline magnitude of \ptmiss, conventionally denoted by \met, is above 200~\GeV.
In addition, events selected by single-electron (muon) triggers~\cite{TRIG-2018-05,TRIG-2018-01} are considered to collect a sample of events to constrain the dominant SM backgrounds with online lepton transverse momentum (\pt) thresholds starting from 24--26 (20--24) \GeV dependent on the period of data taking. The data sample used corresponds to 140~\ifb of $\sqrt{s}=13$~\TeV{} $pp$ collision data, where the uncertainty in the integrated luminosity is 0.83\%~\cite{DAPR-2021-01}. The average number of interactions per bunch-crossing was 33.7.

Samples of Monte Carlo (MC) simulated events are used to estimate the signal yields, and to estimate the background from SM processes with significant~\met{} from invisible particles such as neutrinos.  MC samples are also used to derive systematic uncertainties in the signal and background predictions.

The SUSY signal is modelled using \madgraph~2.7.3~\cite{Alwall:2014hca} with the NNPDF2.3LO~\cite{Ball:2012cx} parton distribution function (PDF) set.  \PYTHIA[8.244]~\cite{Sjostrand:2014zea} is used to model the parton shower, hadronisation, and underlying event, using the A14 set of tuned parameters~\cite{ATL-PHYS-PUB-2014-021}.

Simulated signal samples consist of wino-like \chitz\chiopm, \chiopm\chiomp, \chitz\chitz, and \chiopm\chiopm{} production in association with exactly two additional partons in the final state.\footnote{The \chitz\chitz, \chiop\chiop~and \chiom\chiom~productions feature only pure-electroweak diagrams while the others production modes occur also via mixed electroweak-QCD couplings.} The parton-level jets are required to have $\pt>30~\GeV$ and to be separated by at least 2.5 units in pseudorapidity.  \madspin{}~\cite{Artoisenet:2012st} is used to decay the electroweakinos into a \chioz{} and a pair of SM fermions. The generated samples cover scenarios of $\chitz/\chiopm$ masses between 75--175~\GeV and mass splittings between the NLSPs and LSP of 0.2--5~\GeV. While the signal kinematics do not depend strongly on the electroweakino mass in this regime, samples with larger mass splittings feature more energetic leptons in the decay chain rendering such signals less efficient to the zero-lepton requirement of the analysis. Samples are normalised to leading-order (LO) cross-sections from \madgraph{} assuming pure wino-like states. The total signal cross-section including all production modes ranges from approximately 1.6~pb to 0.06~pb for $\chitz/\chiopm$ masses between 75~\GeV and 175~\GeV. Some kinematics of the SM decay products are sensitive to the relative sign of the signed mass eigenvalues \mchitz{} and \mchioz{}; in this analysis, the product $\mchitz \times \mchioz$ is assumed to be positive.\footnote{The mixing matrix used to diagonalise the neutral electroweakino states can be complex but is forced to be a real matrix in the SLHA2 format~\cite{Allanach:2008qq} at the cost of introducing negative mass eigenstates. The sign will affect the couplings and thus the distributions in the decay under consideration. For additional discussion of this, see Ref.~\cite{Fuks:2017rio} and Appendix A in Ref.~\cite{Gunion:1984yn}.}

Production of two electroweakinos in association with two jets can proceed through either VBF-like diagrams with a $t$-channel process, such as those shown in Figure~\ref{fig:feyn-VBF}, or $s$-channel diagrams such as Figure~\ref{fig:feyn-DY}.  The $s$-channel diagrams can also be either pure-electroweak, in which the boson exchanged by incoming partons is a photon, $W$ boson, or $Z$ boson, or have both electroweak and QCD couplings, in which case the exchanged boson is a gluon.  All such diagrams have identical initial and final states, and all must therefore be included to ensure gauge-invariant predictions.  The interference between these diagrams affects the total cross-section and the kinematics in phase space traditionally targeted by VBF searches, such as large dijet invariant mass ($\mjj$) and large pseudorapidity separation between jets (\detajj).  The effect of the interference in these SUSY models is illustrated in Figure~\ref{fig:interference}, where the impact on the inclusive cross-section is modest, but can be a factor of two or more for $\mjj>5~\TeV$.

\begin{figure}[tbp]
\centering
\includegraphics[width=0.8\columnwidth]{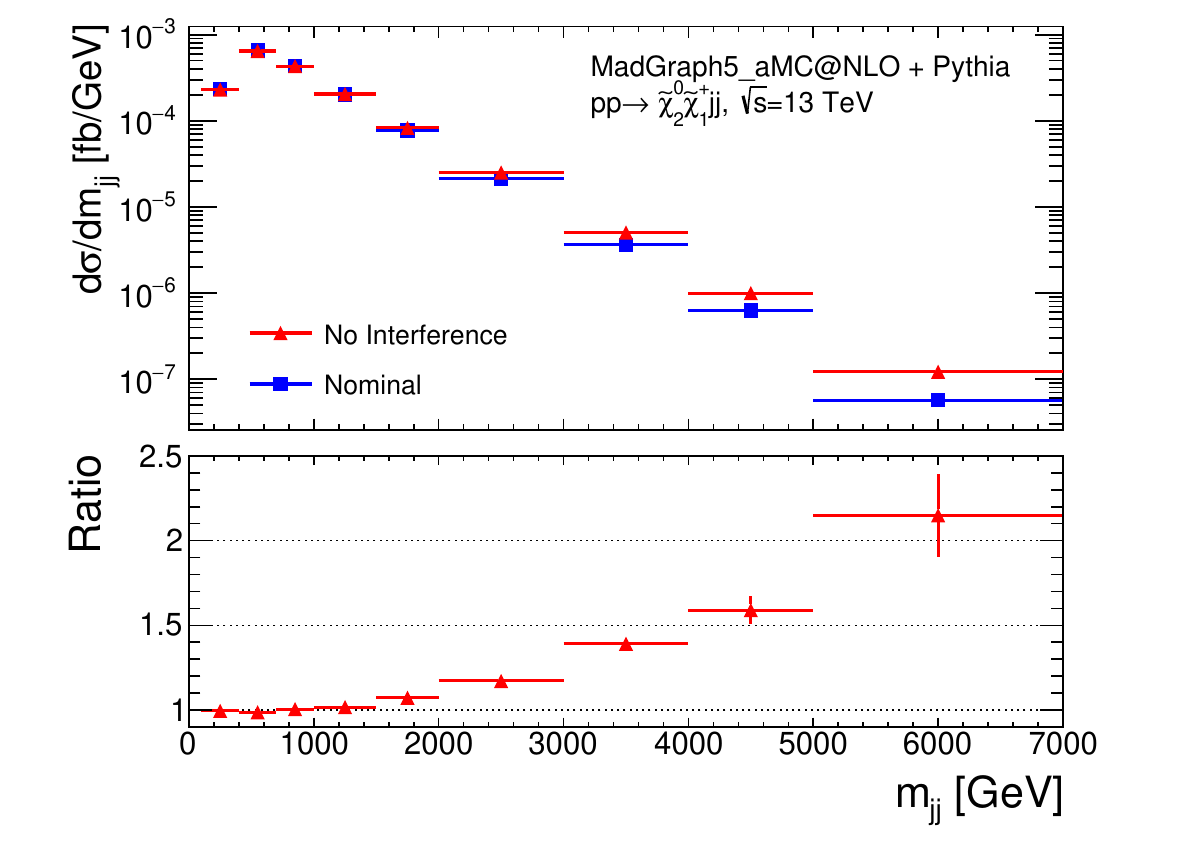}
\caption{Comparison of different process definitions for \chitz\chiop{} production in association with two jets, made at particle level using the anti-$k_t$ algorithm with a radius parameter of $R=0.4$.  The `Nominal' sample is generated according to the details in Section~\ref{sec:samples}.  The `No Interference' sample generates pure-electroweak processes separately from processes with mixed electroweak-QCD couplings.  Events are selected by requiring two jets with $\pt>30~\GeV$ and $|\eta|<5.0$, and oppositely signed pseudorapidities.}
\label{fig:interference}
\end{figure}

The SM background processes are estimated from a combination of MC simulation samples and data-driven approaches.  The most important backgrounds, from SM $Z\to\nu\nu$ and $W\to\ell\nu$ events (where the lepton is not reconstructed), are modelled with \SHERPA~ 2.2.11~\cite{Bothmann:2019yzt} using the NNPDF3.0NNLO PDF set~\cite{Ball:2014uwa}.  For strong production of $V+\text{jets}$ ($V=W,Z,\gamma^*$), where a QCD coupling facilitates the production of the additional jets, processes with up to two coloured partons are modelled at next-to-leading-order (NLO) in the strong coupling, while processes with up to five additional partons are modelled at LO accuracy.  Electroweak production of $V+\text{jets}$ is modelled separately using the same generator, in which processes with up to two partons are modelled at LO accuracy.

Subdominant backgrounds include SM diboson production, events with one or more top quarks, and triboson production.  Diboson production with either fully leptonic or semileptonic decays of the bosons is modelled with \sherpa 2.2.1 or 2.2.2 (depending on the process) at NLO accuracy for up to one additional parton, and at LO accuracy for up to three additional partons. Electroweak diboson production with fully leptonic decays is modelled with \sherpa 2.2.12 at LO accuracy with up to one additional parton. The $t\bar{t}$ and single top-quark processes are generated at NLO with \POWHEGBOX~v2~\cite{Frixione:2007nw,Nason:2004rx,Frixione:2007vw,Alioli:2010xd} for the matrix element, using the NNPDF3.0NLO PDF set~\cite{Ball:2014uwa}, and \PYTHIA[8.230] for the parton shower.  The diagram removal scheme~\cite{Frixione:2008yi} is used to account for interference between $t\bar{t}$ and $Wt$.  Triboson production was modelled with \sherpa 2.2.2 at NLO for inclusive processes, and at LO for up to two additional parton emissions, using the NNPDF3.0NNLO PDF set.

The effect of multiple interactions in the same and neighbouring bunch crossings (\pileup) is modelled by overlaying the simulated hard-scattering event with inelastic $pp$ collisions generated with \PYTHIA[8.186]~\cite{Sjostrand:2007gs} using the NNPDF2.3LO PDF set and the A3 set of tuned parameters~\cite{ATL-PHYS-PUB-2016-017}.  The MC events are weighted to reproduce the distribution of the average number of interactions per bunch crossing observed in the data.

Background and signal samples make use of \EVTGEN 1.6.0 and 1.2.0~\cite{Lange:2001uf} to model the decay of $b$- and $c$-hadrons, with the exception of the background samples modelled with \SHERPA, which performs these decays internally. All MC simulated samples are processed through the ATLAS simulation framework~\cite{SOFT-2010-01} in \GEANT~\cite{Agostinelli:2002hh}, except the signal samples, which are processed with a fast simulation which relies on a parameterisation of the calorimeter response~\cite{ATL-PHYS-PUB-2010-013}.

\section{Event reconstruction}
\label{sec:reco}

Each event is required to have a primary vertex built from at least two associated tracks with $\pT > 0.5~\GeV$. The primary vertex with the highest sum of squared transverse momenta $\sum \pt^2$ of associated tracks~\cite{ATL-PHYS-PUB-2015-026} is selected as the hard-scatter vertex of interest in each event. As described below, reconstructed jets and leptons are categorised according to `baseline' and `signal' criteria, where the latter category includes tighter requirements on the quality of the reconstructed object.

Hadronic jets are reconstructed using the anti-$k_t$ algorithm~\cite{Cacciari:2008gp,Fastjet} with a radius parameter of $R=0.4$ using particle-flow objects~\cite{PERF-2015-09} calibrated at the electromagnetic (EM) scale.  The jet energy scale and resolution are calibrated using simulations with in situ corrections obtained from data~\cite{JETM-2018-05}.  Baseline jets, used for removing overlaps between different types of physics objects, are required to satisfy $\pt > 20~\GeV$ and $|\eta|<4.5$.  Signal jets, which are a subset of baseline jets and are used for event selection and categorisation, must satisfy stricter requirements that suppress contributions from additional $pp$ collisions in the same or nearby bunch crossings. All signal jets are required to have $\pt>30~\GeV$. Central signal jets with $|\eta|<2.4$ must additionally satisfy the \emph{Medium} working point of the jet vertex tagger~\cite{PERF-2014-03} if they have $\pt<60~\GeV$. Signal jets with $2.4<|\eta|<4.5$ must satisfy the \emph{Loose} working point of the forward jet vertex tagger~\cite{ATL-PHYS-PUB-2019-026} if they have $\pT<120~\GeV$. Jets within $|\eta|<2.5$ that satisfy the 85\% efficiency working point of the DL1r algorithm~\cite{FTAG-2019-07} are identified as containing $b$-hadrons and are referred to as $b$-tagged jets.

Baseline electrons are reconstructed using ID tracks matched to energy clusters in the EM calorimeter.  These satisfy $\pt > 4.5$~\GeV{} and $|\eta| < 2.47$ with a \textit{LooseAndBLayerLLH} identification~\cite{EGAM-2018-01}. The longitudinal impact parameter $z_0$ of baseline electron tracks is required to satisfy $|z_0\sin\theta| < 0.5$~mm.  Signal electrons must also satisfy the \textit{Tight} likelihood-based identification criteria as well as the \textit{Loose\_VarRad} isolation requirements~\cite{myEGAM-2021-01}, and have a transverse impact parameter $d_0$ with uncertainty $\sigma(d_0)$ satisfying $\left|d_0/\sigma(d_0)\right| < 5$.

Baseline muons are reconstructed by combining tracks from the ID and the muon spectrometer subsystems. The \textit{Loose} identification criteria~\cite{MUON-2018-03} are applied.  Baseline muons are required to have $\pt > 3~\GeV$ and $|\eta| < 2.7$, and satisfy $|z_0\sin\theta| < 0.5$~mm.  Signal muons must also have $|\eta|<2.5$, and satisfy the \textit{Medium} identification requirements and \textit{Loose$\_$VarRad} isolation criteria.  Signal muons must have impact parameter significance $\left|d_0/\sigma(d_0)\right| < 3$, a tighter requirement than applied for electrons due to the lower levels of material scattering for muons relative to electrons.

To prevent the use of the same reconstructed detector signals in multiple objects, an overlap removal procedure is applied to the baseline leptons and jets in the following order. First, any electron sharing an ID track with a muon is removed. Next, jets are removed if they are within $\Delta R < 0.2$ from a remaining electron. After this, electrons are in turn rejected if they are within $ 0.4 < \Delta R < \min\left(0.4, 0.04 + 10~\GeV / \pt(e)\right)$ of any remaining jet. Jets are removed if they are closer than $\Delta R < 0.4$ to a muon and the jet has fewer than three associated tracks with $\pT > 500 ~\MeV$. Finally, any muon within $ 0.4 < \Delta R < \min\left(0.4, 0.04 + 10~\GeV / \pt(\mu)\right)$ of a jet are removed.

The missing transverse momentum is calculated as the negative vector sum of the transverse momenta of all baseline leptons and jets calibrated to their respective energy scales, and an additional soft term constructed from tracks originating from the hard-scatter vertex but not associated with any of the reconstructed objects~\cite{JETM-2020-03}.

\section{Event selection and analysis strategy}

Several cleaning criteria are applied to all events, including a requirement that the leading jet satisfy the \emph{Tight} jet cleaning criterion to efficiently veto non-collision backgrounds~\cite{DAPR-2012-01}.

All events used in the analysis are required to have zero $b$-tagged jets, and at least two jets with oppositely signed pseudorapidity.  If multiple pairs of jets satisfying that requirement are present in the event, then the jet pair with the largest value of \mjj{} is chosen to represent the VBF tagging jets.  The VBF tagging jet with higher \pt{} is labelled \vbfjetone, while the subleading jet is labelled \vbfjettwo.  Events satisfy the preselection criteria and are retained for further study if they contain a VBF pair that satisfies $\mjj>600~\GeV$, $|\detajj|>3.0$, $\vbfjetonept>80~\GeV$, $\vbfjettwopt>40~\GeV$, and have all baseline jets sufficiently well-separated from the \ptmiss{} in $\phi$ to satisfy $\mindphijetsmet>0.4$ in order to suppress the multijet background.

Events that enter the signal region (SR) are preselected events that first satisfy a set of additional requirements that enhance the sensitivity of the analysis to the SUSY signal.  Specifically, events in the SR must contain zero baseline electrons or muons, satisfy the inclusive-\met{} trigger, and have $\met>250~\GeV$.

A boosted decision tree (BDT) is trained on the preselected signal and SM background MC events described in Section~\ref{sec:samples} to provide the final requirement used to select events for the SR\@. The \textsc{LightGBM}~\cite{10.5555/3294996.3295074} framework is used to construct the BDT, which is trained to separate the SUSY signal from background events originating the SM processes described in  Section~\ref{sec:samples}, weighted according to their cross-section. Gradient boosting with the binary cross entropy as optimisation objective was used to train the BDT\@. The maximum number of trees in the BDT is restricted to 1500 where each tree has a maximum depth and number of leaves of five and eight, respectively. Only signal samples with $m($\chitz,\chiopm$) = 100~\GeV$ were supplied in the training to optimise the BDT performance for this mass regime. A $k$-fold cross-validation (with $k=5$) approach is employed, with one fold used for validation, one fold used for testing, and the remaining folds used for training.

The list of input variables for the BDT consists of:
\met;
\pt of the two individual VBF-tagged jets;
the \pt sum of the VBF-tagged jets as well as the same sum divided by \met;
the separation in pseudorapidity \detajj, and separation in azimuthal angle \dphijj, of the two VBF-tagged jets;
the separation in $\phi$ between each of the two VBF tagged jets and \ptmiss, $\Delta\phi(\mathbf{j},\ptmiss)$, the absolute value of the difference $|\Delta\phi(\vbfjone,\ptmiss)-\Delta\phi(\vbfjtwo,\ptmiss)|$, and the $\phi$ separation between \ptmiss and the vector sum of the two VBF tagged jets;
two transverse masses \mt, one for each of the VBF-tagged jets, as well as the sum of the two transverse masses, where the transverse mass is computed as $\mt(\mathbf{j},\ptmiss)=\sqrt{2\pt^{j}\met(1-\cos(\Delta\phi(\mathbf{j},\ptmiss)))}$;
properties of the VBF jet system (\mjj{} and \pt);
the weighted sum of \mjj{} and \met, where the weighted sum is calculated as \mjj/1430~\GeV + \met/320~\GeV with the `normalisation' values in the denominators representing the average of \mjj and \met of the SM background at preselection level, respectively;
the minimum $\phi$ separation of any baseline jet with the \ptmiss;
and the \mt{} and separation in $\phi$ of the $\Upsilon$- and $\Xi$-tagged jets with the \ptmiss, where jet $\Upsilon$ minimises $\Delta\phi(\mathbf{j},\ptmiss)$, and jet $\Xi$ minimises $\mt(\mathbf{j},\ptmiss)$ using all baseline jets passing the overlap removal as candidates for $\Upsilon$ and $\Xi$.

The result of the BDT is a score assigned to each event, where scores close to unity are more consistent with signal processes, while scores close to zero are more similar to background processes. The most important input features were found to be \detajj and the \pt of the subleading jet. Only events with scores above 0.6 are kept in the SR; control and validation regions use events with BDT scores as low as 0.4, as described in Section~\ref{sec:backgrounds}.  Depending the region, the BDT score distribution is either considered inclusively within some range, or binned in steps of 0.04, except the last bin, which covers scores from 0.88 to 1.0. This last bin shows the best sensitivity to the benchmark signals and is hence also used as single-bin SR to derive constraints on the number of events from generic physics processes beyond the SM as described in Section~\ref{sec:result}. After the training of the BDT, events in the SR are further split into two channels based on their jet multiplicity ($\njets$): one channel containing events with exactly two jets (dubbed as `\twoj' in the following), and another containing events with three or more jets (`\threej'). This split mildly increases the sensitivity but in particular takes into account the observed dependence of the normalisation and modelling uncertainties of the \Vjets backgrounds as described below.

\section{Background estimation}
\label{sec:backgrounds}

Predictions for SM background contributions to the SR are made using different strategies depending on the background process.  The largest backgrounds arise from $W$ and $Z$ bosons produced in association with two or more jets, which are modelled using MC samples that are normalised in dedicated control regions (CRs).  Backgrounds from processes that produce \met{} through jet mismeasurement or instrumental effects (referred to here as multijet backgrounds) are modelled using a data-driven method.  All other subdominant backgrounds listed in Section~\ref{sec:samples} are modelled using MC.

\begin{table}[tbp]
\begingroup
\caption{Summary of the selections defining the CRs, VRs and SRs. Each region is additionally split into \twoj and \threej categories. The requirements of the preselection described in the text and on the number of baseline and signals leptons defined in Section~\ref{sec:reco} are also applied. The $\met/\sqrt{\Sigma E_{\mathrm{T}}}$ requirement for the $W$ + jets regions is only applied to events with electrons, to enhance the fraction of $W$ + jets in those regions.}
\label{tab:bkgs:regions}
\renewcommand{\arraystretch}{1.25}
\maxsizebox*{\textwidth}{\textheight}{
\begin{tabular}{l @{\extracolsep{2em}}c@{\extracolsep{0em}}c @{\extracolsep{2em}}c@{\extracolsep{0em}}c  @{\extracolsep{2em}}c@{\extracolsep{1em}}c@{\extracolsep{1em}}c}
\toprule
\toprule
Feature & \CRZ & \VRZ & \CRW & \VRW & \VRZero & Multi-bin \SR  & Single-bin SR \\
\cmidrule(r){1-1}\cmidrule(lr){2-3}\cmidrule(lr){4-5}\cmidrule(l){6-8}
$N_{\mathrm{leptons}}$ & \multicolumn{2}{c}{2} & \multicolumn{2}{c}{1} & & 0 &\\%
$m_{\ell\ell}$  &\multicolumn{2}{c}{$|m_{\ell\ell} - m_{Z}| < 30~\GeV$}  &\multicolumn{2}{c}{--} &    & -- &    \\
$\met/\sqrt{\Sigma E_{\mathrm{T}}}$  &\multicolumn{2}{c}{--}  &\multicolumn{2}{c}{$\met/\sqrt{\Sigma E_{\mathrm{T}}} > 5~\sqrt{\GeV}$} &   & -- &   \\
BDT score & [0.50, 0.84) & [0.84, 1.0] & [0.50, 0.84) & [0.84, 1.0] & [0.4, 0.6) & [0.6, 1.0] & [0.88, 1.0] \\
BDT score bins  & 1 & 2 & 1 & 2 & 5 & 8 & 1\\
\bottomrule
\bottomrule
\end{tabular}
}
\endgroup
\end{table}


Monte Carlo predictions of $Z+\text{jets}$, where the $Z$ boson decays into $\nu\bar{\nu}$, are normalised in CRs identical to the SRs except that they require the presence of two opposite-sign, same-flavour signal leptons with an invariant mass within 30~\GeV{} of the $Z$-boson mass.  The lepton momenta are added to the \ptmiss{} for the purposes of applying the \met{} requirements in the SR and calculating the BDT inputs using this adjusted definition of \ptmiss. This approach allows using $Z\to ee/\mu\mu$ events as proxy for the $Z\to\nu\bar{\nu}$ background present in the SR\@. Events in \CRZ are collected using single-lepton triggers. To ensure stable trigger efficiencies the leptons in \CRZ are required to be sufficiently energetic and lie within detector acceptance. Thus, the leading lepton is required to have a $\pt > 27~\GeV$ and must be matched to the online trigger object. To efficiently veto mis-identified electrons, the energy deposits of each electron in the EM calorimeter must satisfy $|\eta|<1.37$ or $1.52<|\eta|<2.47$ and the subleading electron must have $\pt > 25~\GeV$. Subleading muons must have $\pt>9~\GeV$, with both muons satisfying $|\eta|<2.5$.

The $W+\text{jets}$ predictions are normalised in separate CRs that includes events with exactly one signal lepton.  As with \CRZ, the lepton momentum is added to the \ptmiss{} before application of the remaining SR criteria and calculation of the BDT score. A combination of single-lepton triggers and the inclusive-\met{} trigger are used to collect events in this CR\@.  Due to larger rates of multijet processes in which a hadron (or one of its decay products) is misidentified as a lepton, additional kinematic requirements are imposed to enhance the fraction of $W+\text{jets}$ in this region.  This includes requiring leptons to satisfy $\pt>27~\GeV$, and for electron events requiring $\met/\sqrt{\Sigma E_{\mathrm{T}}} >5.0~\sqrt{\GeV}$, where $\Sigma E_{\mathrm{T}} = \pt(j_1)+\pt(j_2)+\pt(e)$, with $\met/\sqrt{\Sigma E_{\mathrm{T}}}$ used as an approximation of the \met{} significance~\cite{ATLAS-CONF-2018-038}.

Events in \CRW and \CRZ are required to have BDT scores between 0.5 and 0.84, and are split into \twoj and \threej channels in accordance with the SRs. This allows the normalisation of the $W$ and $Z$ backgrounds individually for each category, resulting in four normalisation factors in total. While the lower requirement on the BDT score ensures that the normalisation factors are derived in a phase space similar to the SRs, events with BDT scores between 0.84 and 1.0 are reserved for validation regions (VRs) used to check the extrapolation of the $V+\text{jets}$ predictions into high-BDT regions, which provide the best sensitivity to SUSY signals in the SRs. Distributions of the BDT scores for \CRW and \CRZ in both the \twoj and \threej channels are shown in Figure~\ref{fig:WZCR}. In addition, a separate set of validation regions, called VR-0L, are constructed from SR events with BDT scores between 0.4 and 0.6 to validate the extrapolation of the $V$ + jets normalisations derived in one- or two-lepton regions to the zero-lepton phase space. A summary of the selections employed to define the regions used in the analysis is presented in Table~\ref{tab:bkgs:regions}.

\begin{figure}[tbp]
\subfloat[]{
\includegraphics[width=0.48\columnwidth]{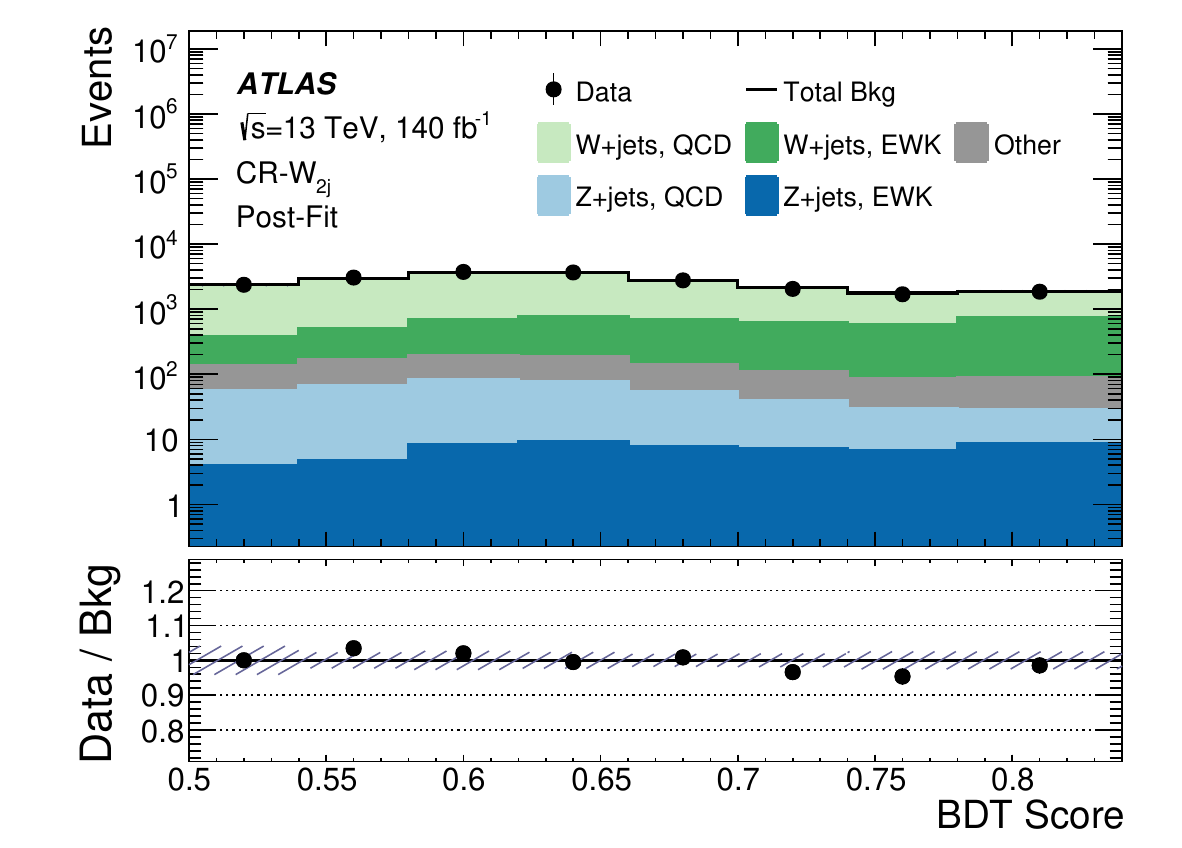}
}
\subfloat[]{
\includegraphics[width=0.48\columnwidth]{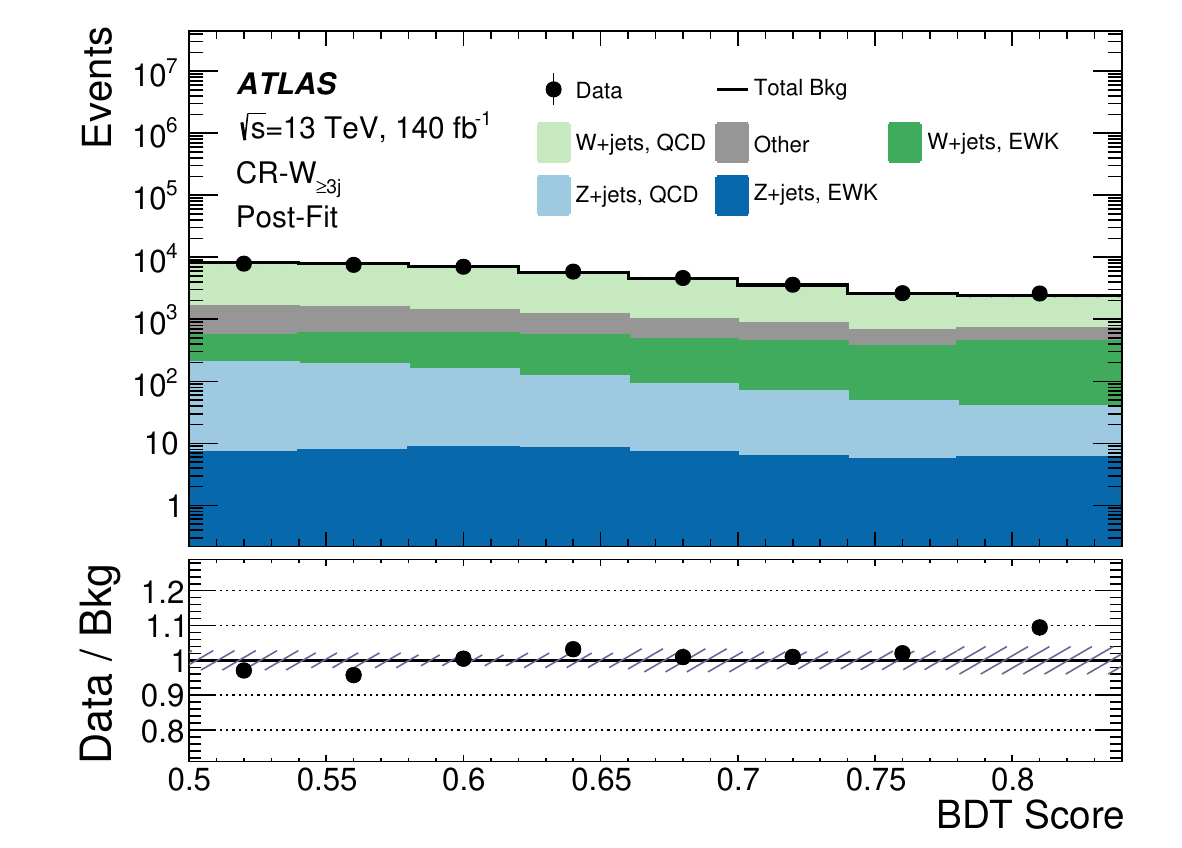}
}\\
\subfloat[]{
\includegraphics[width=0.48\columnwidth]{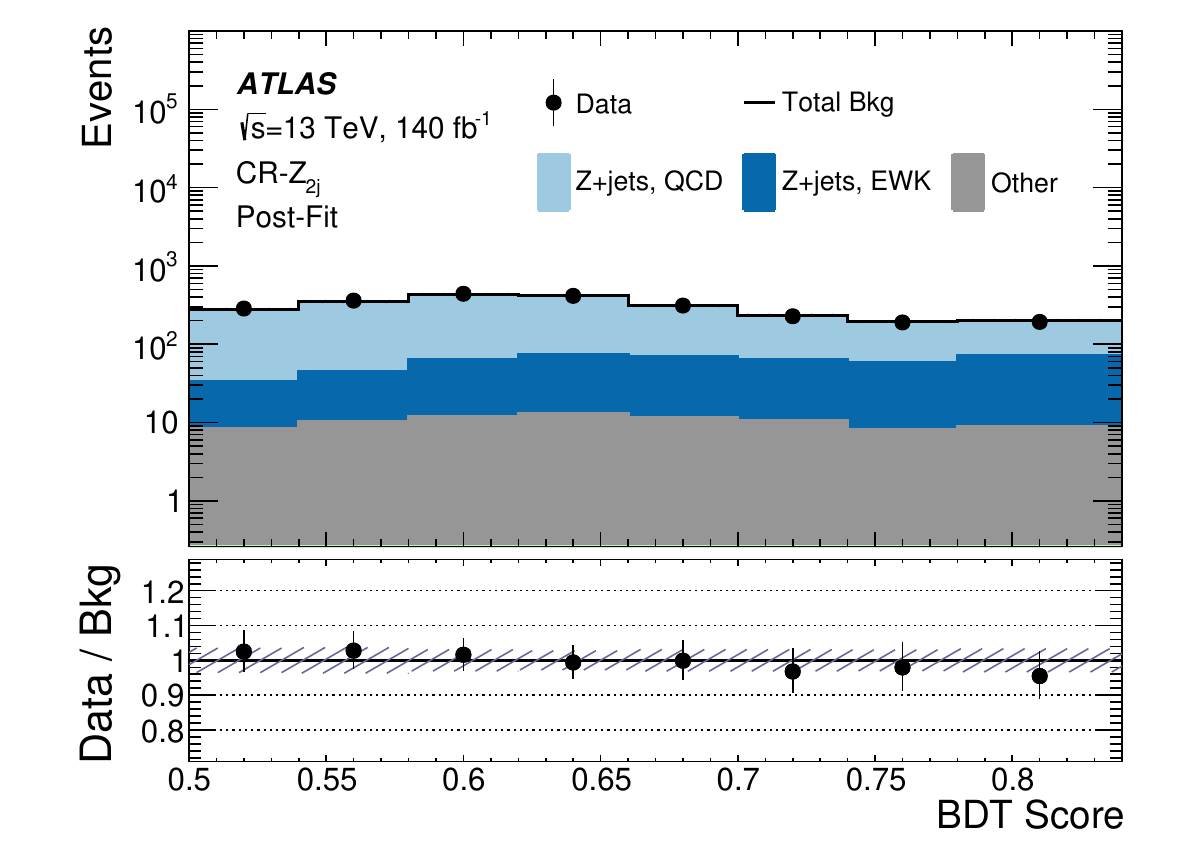}
}
\subfloat[]{
\includegraphics[width=0.48\columnwidth]{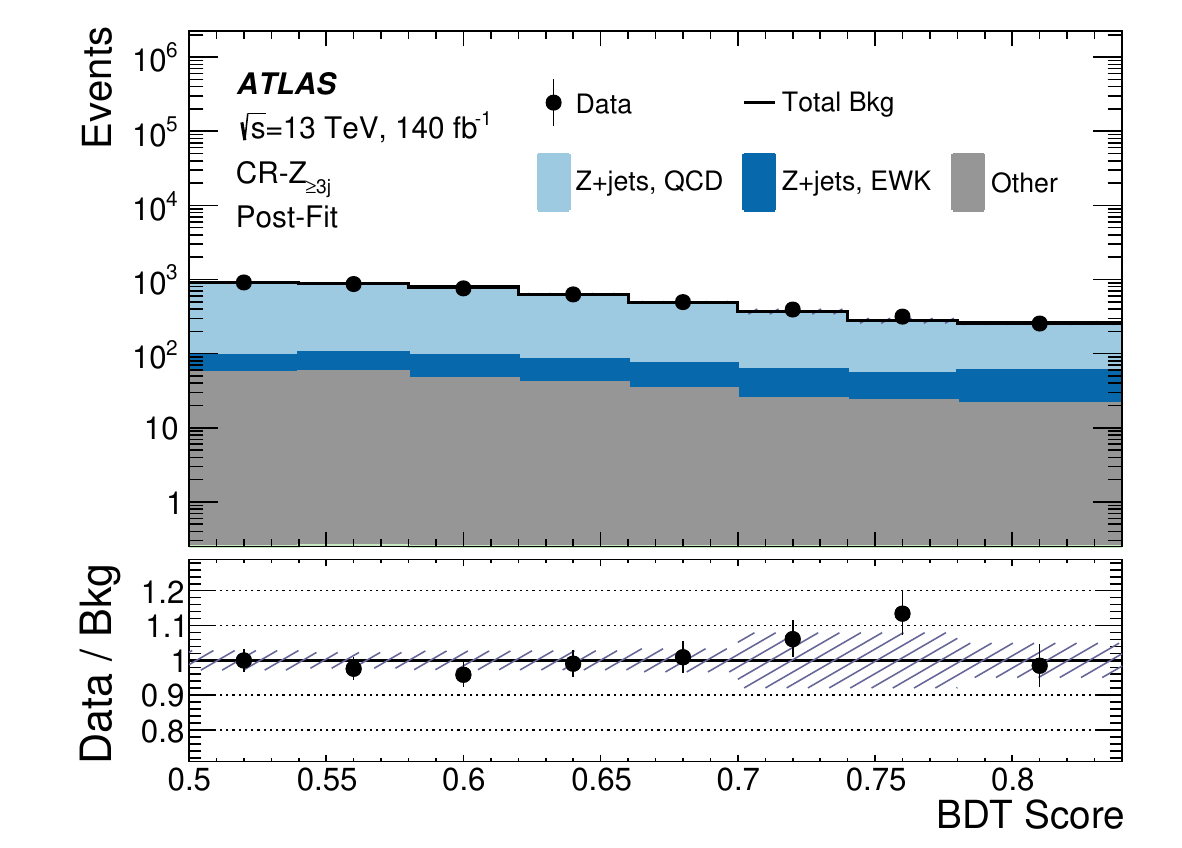}
}
\caption{The BDT score distribution in (a) \CRWtwoj, (b) \CRWthreej, (c) \CRZtwoj, and (d) \CRZthreej{} used to normalise the $W+\text{jets}$ and $Z+\text{jets}$ estimates. The total background prediction is shown after the fit to the data. The `Other' category contains rare backgrounds from diboson, triboson and top-quark production processes. The hatched band represents the post-fit experimental, theoretical, and statistical uncertainties in the total background.  The bottom panel of each plot shows the ratio between the data and the post-fit background prediction.}
\label{fig:WZCR}
\end{figure}

A small but non-negligible amount of SM background also originates from multijet processes in which the \met originates from mismeasured jets. It would be computationally expensive to simulate this background with sufficient MC statistics in the high-\met phase space considered in this search. Hence, multijet backgrounds are estimated via a semi data-driven approach using dedicated CRs enriched by multijet contributions utilising an `ABCD'-like region layout where `D' labels the phase space of the SR for which the multijet estimate is derived. The region CR-A is identical to the SR, except for a requirement that $\mindphijetsmet<0.4$.  Regions CR-B and CR-C are identical to CR-A and the SR, respectively, except that they require events to have $200 < \met/\GeV < 220$ and BDT scores above 0.4 to be orthogonal but still kinematically similar to the SR\@.  The inclusive yield $N_{\textrm{SR}}$ of multijet events in the SR, i.e. integrated over all SR bins, is then computed as

\begin{equation}
N_{\textrm{SR}} = \frac{N_{\textrm{CR-C}}^{\textrm{data}}-N_{\textrm{CR-C}}^{\textrm{MC}}}{N_{\textrm{CR-B}}^{\textrm{data}}-N_{\textrm{CR-B}}^{\textrm{MC}}}\left(N_{\textrm{CR-A}}^{\textrm{data}}-N_{\textrm{CR-A}}^{\textrm{MC}}\right)\;,
\end{equation}

where e.g. $N_{\textrm{CR-C}}^{\textrm{data}}$ represents the observed data yield in CR-C. Contributions from other SM processes to all three CRs are estimated with MC and subtracted from the data yields, denoted by e.g. $N_{\textrm{CR-C}}^{\textrm{MC}}$.  A minor dependence of the $V+\text{jets}$ normalisation with respect to \mindphijetsmet~was found, thus predictions from $V+\text{jets}$ are scaled using dedicated one- and two-lepton CRs for which $\mindphijetsmet<0.4$ before performing the MC subtraction. Except for the lepton requirements, these CRs are otherwise identical to CR-A/B.  The obtained multijet predictions in CR-A are binned as a function of the BDT score, using identical binning as in the SR\@.  These vary smoothly as a function of the BDT score, dropping to near zero for BDT scores above 0.84. To mitigate the impact of statistical fluctuations, a $\chi^2$-based fit is performed on the binned multijet estimates using the BDT score as a dependent variable. Several functions were tested to fit the spectrum where a parameterisation via a five-parameter `Gaussian + linear' function yielded the best results. The final multijet estimates are then taken as the values of the fitted function at the centres of each bin. Tests of this procedure in a region similar to the SR but requiring $220 < \met/\GeV < 250$ show that this data-driven estimate significantly improves the comparison between predictions and data relative to multijet estimates taken from MC where the latter also suffer from large statistical fluctuations.

\section{Systematic uncertainties}

Predictions of the BDT score distribution in the signal regions are affected by systematic uncertainties, which are due to both experimental effects and MC generator uncertainties.  The $V+\text{jets}$ backgrounds are normalised in dedicated control regions such that systematic uncertainties in those backgrounds only affect the extrapolation of the $V+\text{jets}$ predictions from the control regions to the signal regions.  A summary of all systematic effects is shown in Figure~\ref{fig:systematics}.

\begin{figure}[tbp]
\includegraphics[width=\columnwidth]{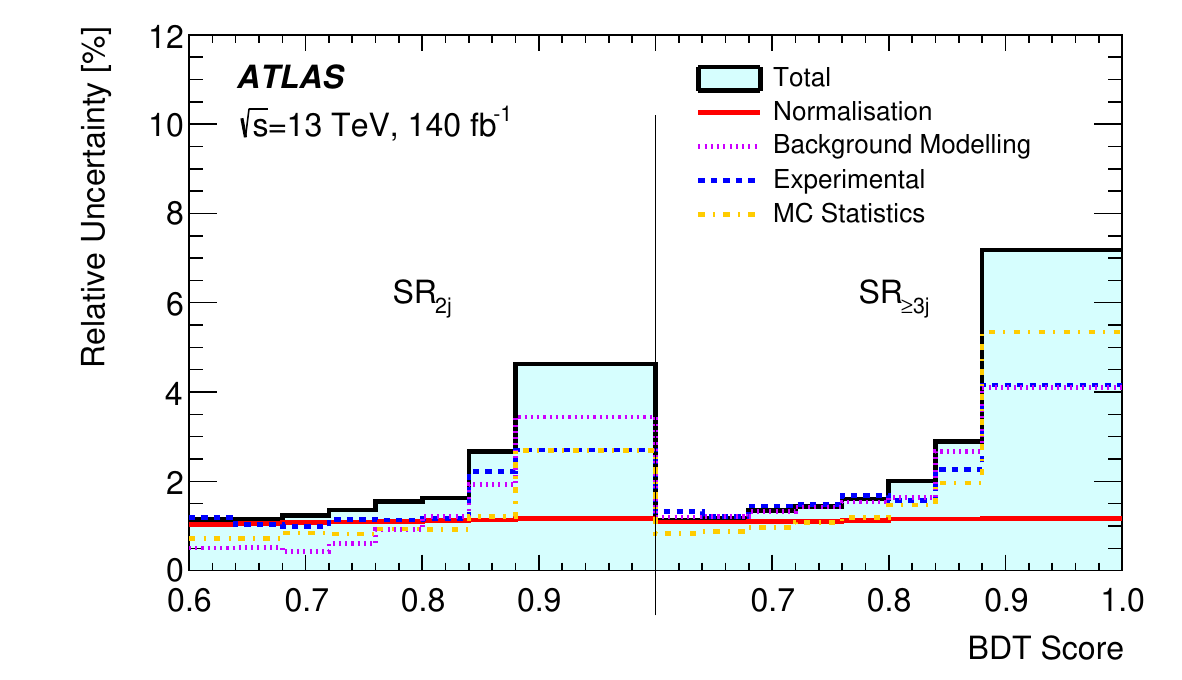}
\caption{The relative systematic uncertainties in the BDT score distribution for background after the exclusion fit. The `Normalisation' uncertainty arises from the use of CRs to normalise $W$ + jets and $Z$ + jets, while `Background Modelling' includes the different sources of theoretical modelling uncertainties in the distribution of BDT scores, as well as the uncertainties in the data-driven multijet background estimate. The uncertainties associated with the measurement of the integrated luminosity as well as the reconstruction and selection of signal leptons, jets and \met{} are included under the `Experimental' category. The `MC Statistics' uncertainty originates from the limited size of the MC samples used to model the irreducible background contributions. The individual uncertainties can be correlated and do not necessarily add up in quadrature to the total uncertainty.}
\label{fig:systematics}
\end{figure}

The trigger, reconstruction, identification and isolation efficiencies for electrons~\cite{EGAM-2018-01} and muons~\cite{PERF-2015-10}, as well as the momentum resolution and scale, primarily affect the signal region through \CRW and \CRZ.  The uncertainty in the muon reconstruction efficiency was found to be the largest experimental effect on the background yield in the SR as it restricts how precisely the normalisation of the \Vjets backgrounds can be determined.

Other sources of experimental systematic uncertainties are the jet energy scale and resolution~\cite{PERF-2016-04}.  The systematic uncertainties related to the modelling of \met\ in the simulation are estimated by propagating the uncertainties in the energy and momentum scale of each of the objects entering the calculation, as well as the uncertainties in the soft-term resolution and scale~\cite{JETM-2020-03}.

Several uncertainties arise from MC generator modelling of SM backgrounds.  The uncertainties related to the choice of the QCD renormalisation and factorisation scales are assessed by varying the corresponding generator parameters in the matrix element and parton shower up and down by a factor of two around their nominal values, removing combinations where the variations differ by a factor of four. The resulting yield variations for $V+\text{jets}$ are approximately 20\% for both strong and electroweak $V+\text{jets}$ samples before normalisation in the control regions, except for strong $V+\text{jets}$ in the \twoj regions where variations range from 25\% to 45\%. An additional uncertainty arises from the scheme, which can be additive, multiplicative, or exponential, used to evaluate NLO electroweak corrections for the strong $V+\text{jets}$ backgrounds. The uncertainty in this choice is calculated as the difference in predictions between the scheme that is closest to the nominal and that which deviates most from nominal. The uncertainties associated with the choice of PDF set, NNPDF~\cite{Butterworth:2015oua,Ball:2012cx,Ball:2014uwa}, and the uncertainty in the strong coupling constant, $\alpha_\mathrm{s}$, used in the PDF evaluation are also included. These modelling uncertainties are treated as uncorrelated  between the $\Wjets$ and $\Zjets$ backgrounds as well as between their strong and electroweak contributions. To take into account imperfect modelling of the minor backgrounds such as top-quark production, a normalisation uncertainty of 30\% is added to these in the SRs. The impact of these normalisation uncertainties were found to be below 2\% on the final results.

Theoretical uncertainties in the signal predictions include uncertainties in the cross-section and shape uncertainties due to the renormalisation and factorisation scales, choice of PDF set and parton showering. The total cross-section uncertainty was found to be of around 10\% and 25\% for pure-electroweak and mixed electroweak-QCD production modes at electroweakino mass of 100~\GeV, respectively.

Several sources of uncertainty are considered for the multijet background estimate and implemented as individual nuisance parameters.  Validation regions with $220 < \met/\GeV < 250$, i.e. adjacent to CR-A, CR-B, CR-C and CR-D, are used to estimate an uncertainty in the extrapolation from the CR to the SR phase space that is performed in the estimation. The fitted parameters of the Gaussian + linear fit to the BDT distribution in CR-A are varied within their 1$\sigma$ confidence intervals to generate alternative estimates of the multijet background; the variation leading to the largest deviation of the multijet prediction is used to define the fit uncertainty.  The impacts of uncertainties on leptons, jets, and \met{} as well as theoretical uncertainties in the MC predictions in CR-A, CR-B, CR-C, and CR-D are also considered. As CR-B and CR-C are located within a \met regime where the \met trigger might not be fully efficient yet dependent on the event topology, a 5\% uncertainty is considered to cover any potential mismodelling of the trigger efficiency in MC simulation.
The total pre-fit uncertainty in the multijet background estimate is found to be around 100\% in the inclusive \SRtwoj and \SRthreej.

As seen in Figure~\ref{fig:systematics}, there is no clear dominant source of systematic uncertainty among any of the SR bins; rather, all sources contribute roughly equally to the total uncertainty in the background estimates.

\section{Results}
\label{sec:result}

Data in the CRs, VRs and SRs are compared with SM predictions using a profile
likelihood method~\cite{Cowan:2010js} implemented in the \textsc{HistFitter} package~\cite{Baak:2014wma}.
Systematic uncertainties are treated as nuisance parameters with Gaussian constraints in the likelihood, where experimental systematic uncertainties are correlated between signal and backgrounds for all regions.

A background-only fit is performed using only \CRW and \CRZ to constrain the normalisation parameters of the \Wjets and \Zjets backgrounds.  The resulting normalisation parameters are shown in Table~\ref{tab:normfactors}.  With these normalisation parameters applied, the background prediction is checked in \VRW, \VRZ and VR-0L, as shown in Figures~\ref{fig:VRWZ} and~\ref{fig:VR0L}, respectively. All VRs are shown separately in the \twoj and \threej channels. Overall, the predictions agree well with the observed data within the systematic uncertainties, confirming the applicability of the derived \Vjets normalisations at large BDT scores and to zero-lepton events. The second bin of \VRWtwoj shows a small excess that is slightly below the $2\sigma$ level, which is interpreted as a statistical fluctuation of the data. Overall, the deviations between the background predictions and the observed yields in all VRs are within two standard deviations.

\begin{table}[tbp]
\centering
\caption{Normalisation factors for $W$ + jets and $Z$ + jets MC samples derived from one- and two-lepton control regions. The uncertainties include both statistical and systematic contributions.}
\renewcommand{\arraystretch}{1.1}
\begin{tabular}{r c c}
\toprule
\toprule
Region   & \CRW   & \CRZ \\
\midrule
\twoj     &0.800 $\pm$ 0.007           &0.829 $\pm$ 0.018         \\
\threej   &0.986 $\pm$ 0.013           &1.046 $\pm$ 0.019         \\
\bottomrule
\bottomrule
\end{tabular}
\label{tab:normfactors}
\end{table}


\begin{figure}[!htb]
\centering
\subfloat[]{
\includegraphics[width=0.48\textwidth]{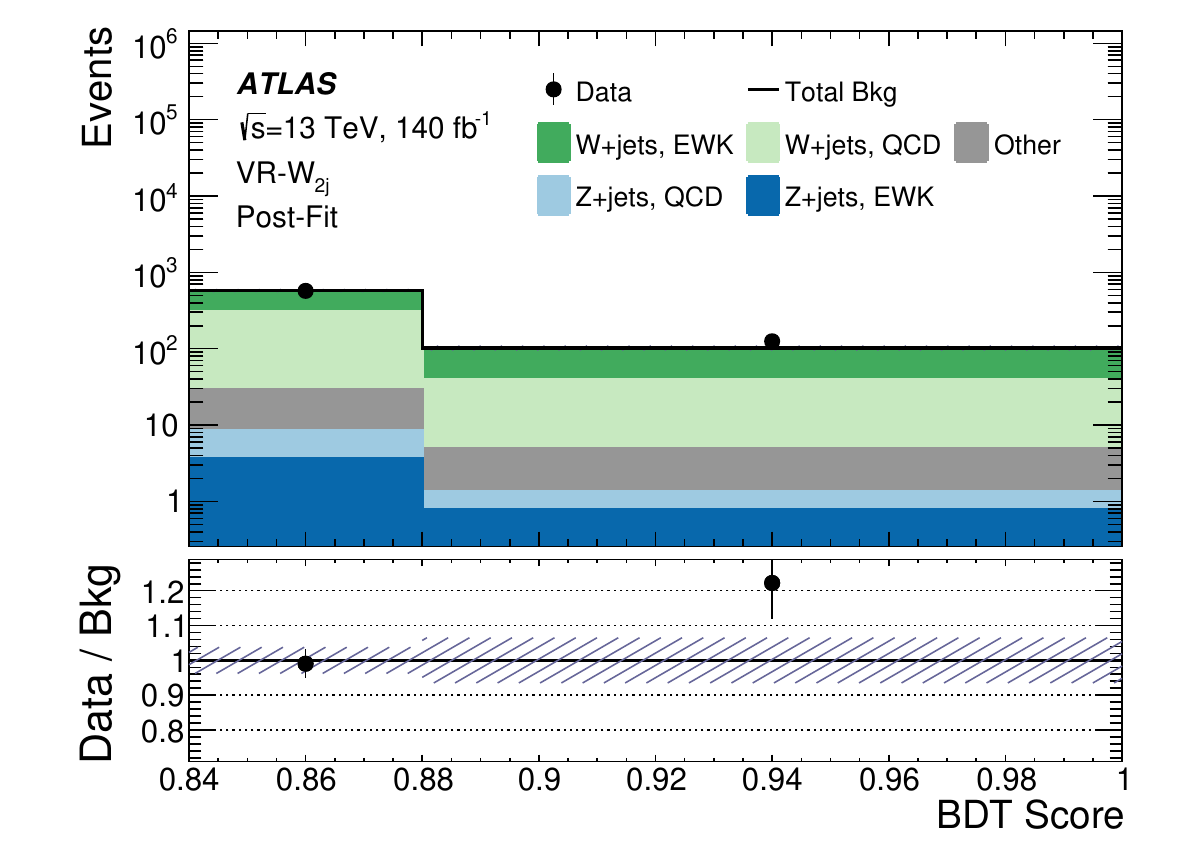}
}
\subfloat[]{
\includegraphics[width=0.48\textwidth]{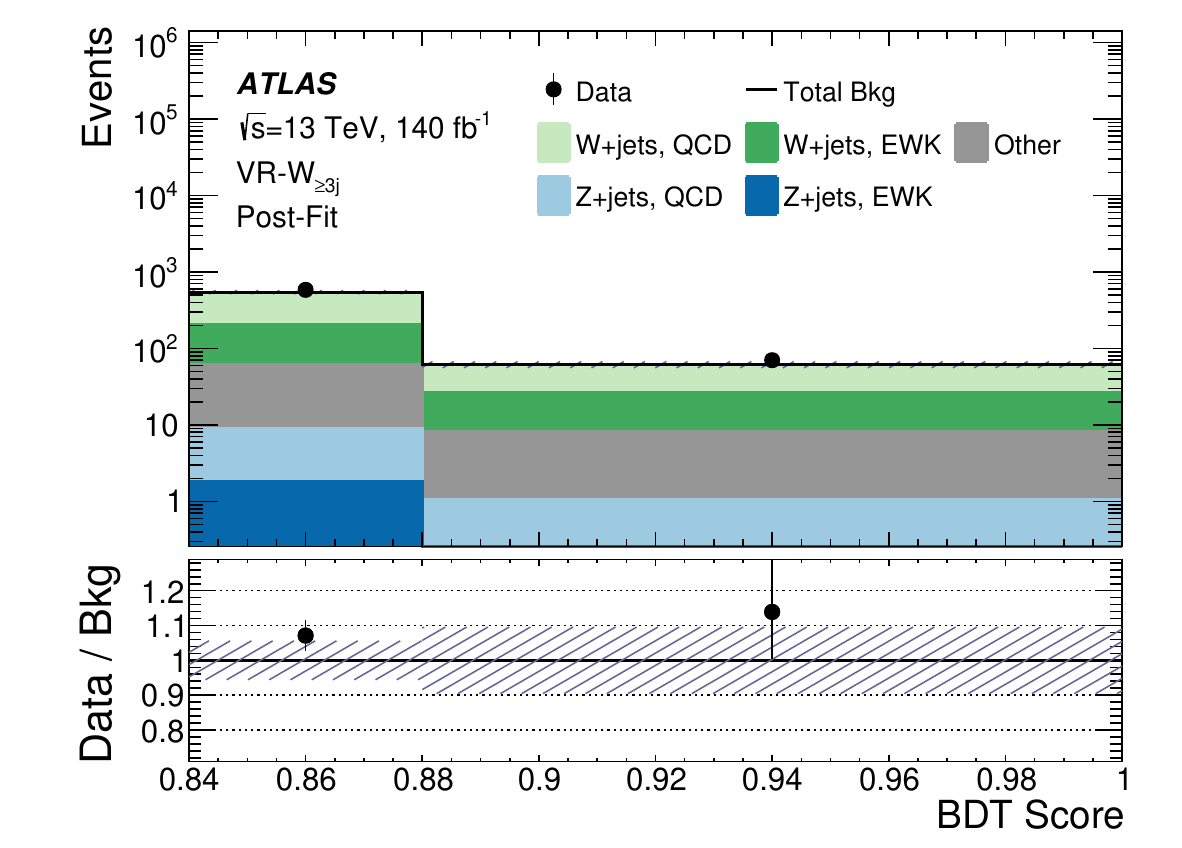}
}\\
\subfloat[]{
\includegraphics[width=0.48\textwidth]{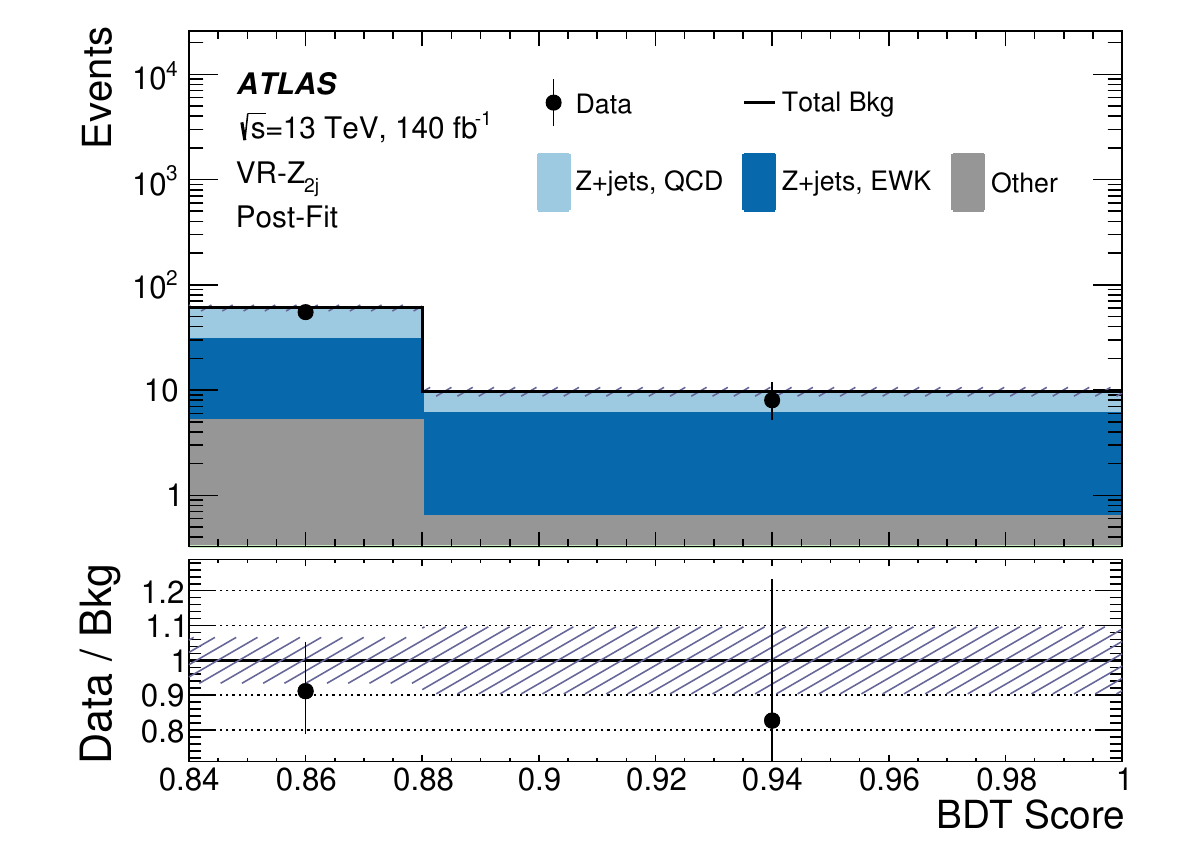}
}
\subfloat[]{
\includegraphics[width=0.48\textwidth]{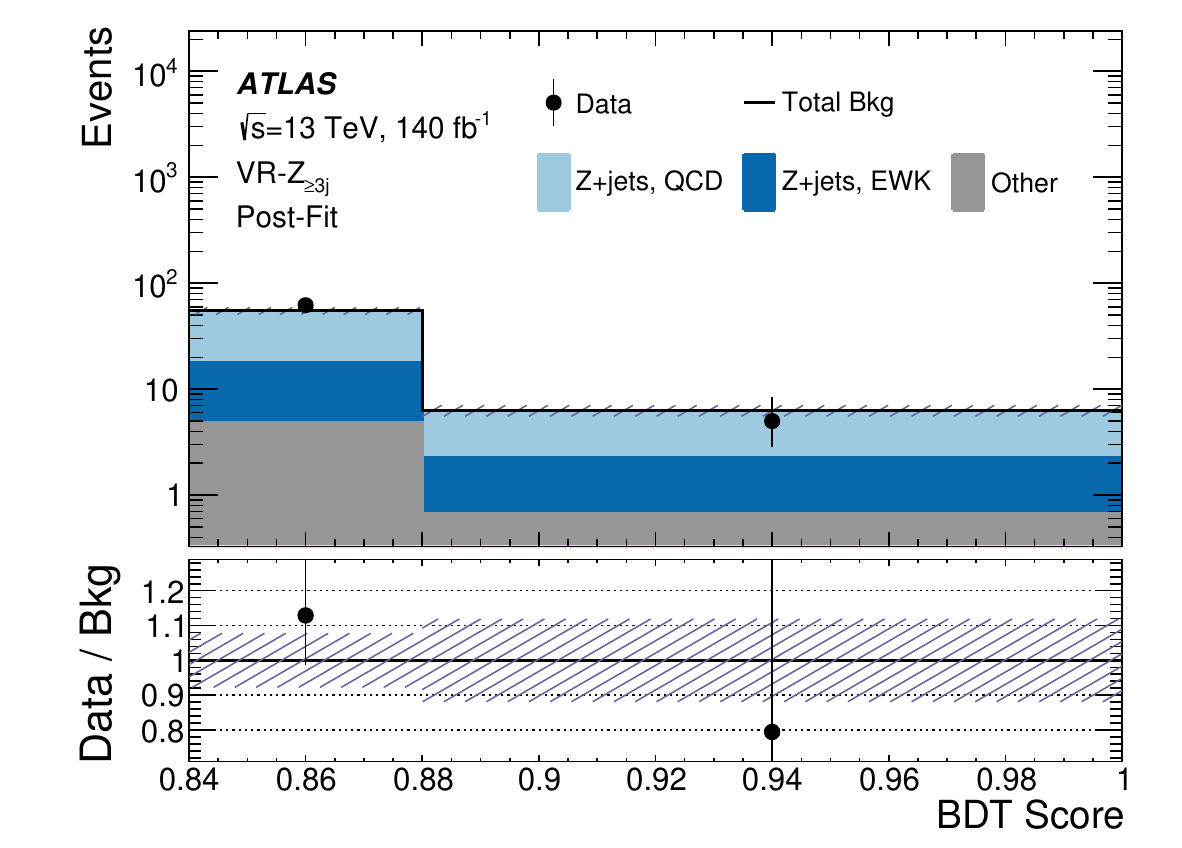}
}
\caption{Observed and predicted background distributions of the BDT score in (a) \VRWtwoj, (b) \VRWthreej, (c) \VRZtwoj, and (d) \VRZthreej{} validation regions shown after the control region fit. The `Other' category contains rare backgrounds from diboson, triboson and top-quark production processes. The hatched band represents the post-fit experimental, theoretical, and statistical uncertainties in the total background. The bottom panel of each plot shows the ratio between the data and the post-fit background prediction.}
\label{fig:VRWZ}
\end{figure}

\begin{figure}[!htb]
\centering
\subfloat[]{
\includegraphics[width=0.48\textwidth]{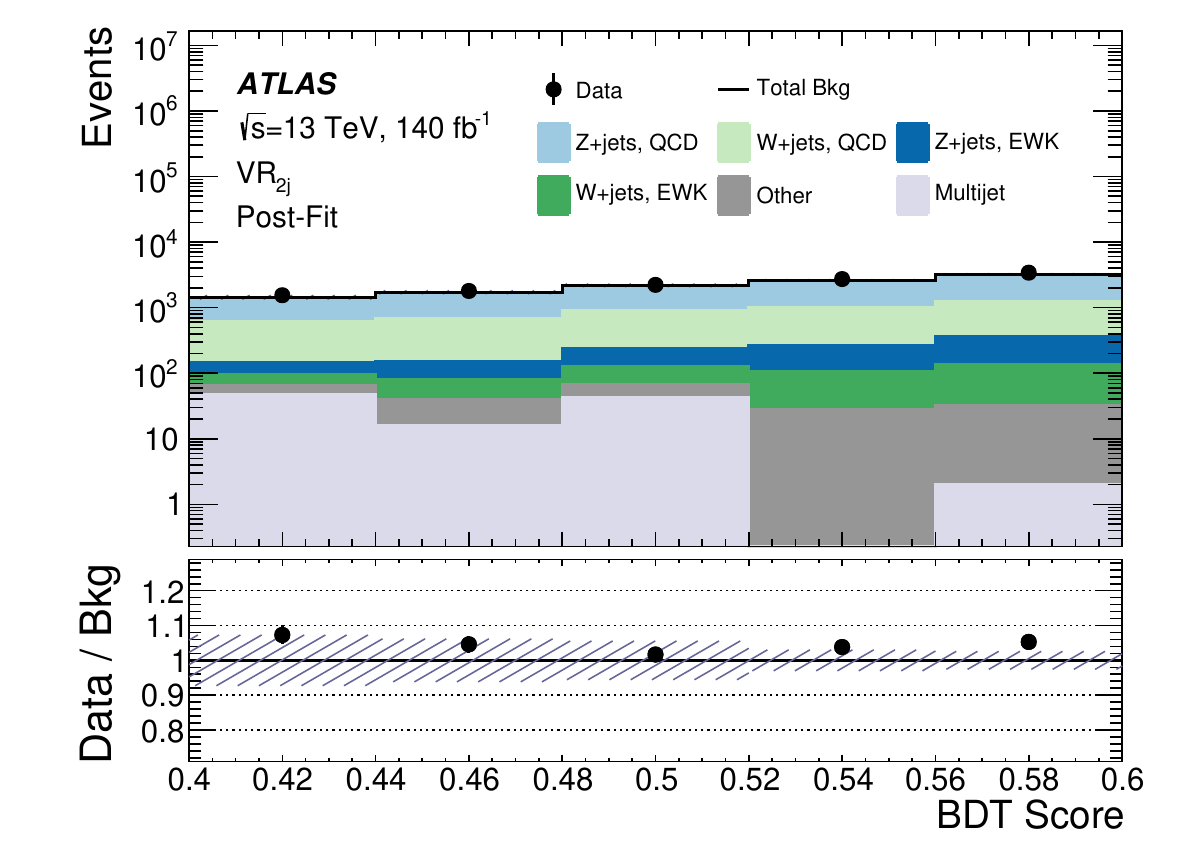}
}
\subfloat[]{
\includegraphics[width=0.48\textwidth]{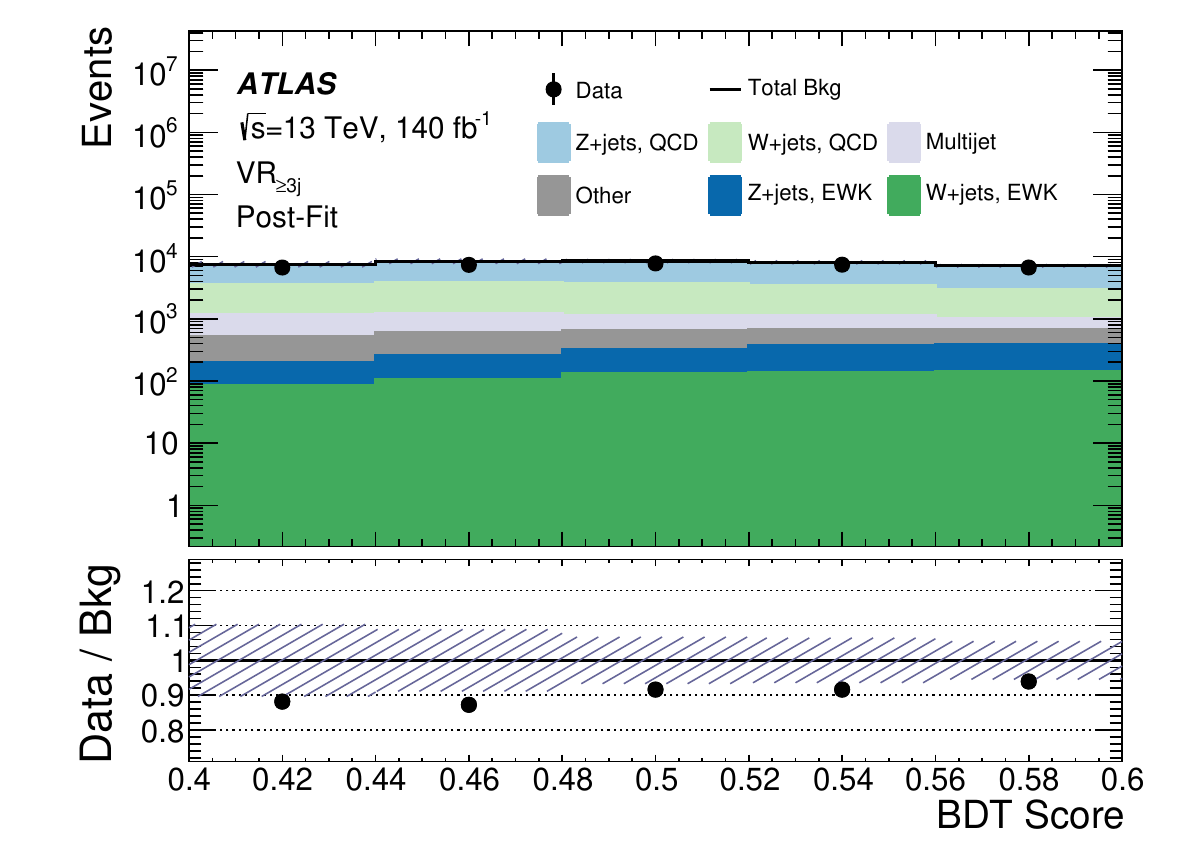}
}
\caption{Observed and predicted background distributions of the BDT score in (a) \VRZeroLtwoj{} and (b) \VRZeroLthreej low-BDT validation regions, using data-driven estimates for the multijet background. The `Other' category contains rare backgrounds from diboson, triboson and top-quark production processes. The hatched band represents the post-fit experimental, theoretical, and statistical uncertainties in the total background.  The bottom panel of each plot shows the ratio between the data and the post-fit background prediction.}
\label{fig:VR0L}
\end{figure}

To test for the presence of excesses, the background-only fit is extended to include a single-bin signal region consisting of events in either \SRtwoj{} and \SRthreej{} with BDT scores greater than 0.88. An independent fit is performed in each single-bin region containing a signal model with an unconstrained normalisation parameter to estimate the contributions of any phenomena beyond those predicted by the SM\@.  Each region is fit simultaneously with the control regions, which are assumed to contain no signal.  To quantify the probability under the background-only hypothesis to produce event yields greater than or equal to the observed data, $p$-values are calculated for each single-bin region. The results of the two fits are shown in Table~\ref{tab:discovery}.  The lowest $p$-value is observed in the \threej{} region, corresponding to a significance of less than $1\sigma$.  The CL$_\text{s}$ prescription~\cite{Read:2002hq} is used to perform a hypothesis test that sets upper limits at the 95\% confidence level (CL) on the observed (expected) number of signal events $S_\text{obs (exp)}^{95}$ in each single-bin region.  Dividing $S_\text{obs}^{95}$ by the integrated luminosity defines the upper limits on the visible cross-sections $\langle\epsilon\sigma\rangle_\text{obs}^{95}$. Generic non-SM processes that predict more than 13 (25) events in \SRtwoj{} (\SRthreej{}) with BDT scores greater than 0.88 are excluded at 95\% CL, with corresponding upper limits on their visible cross-sections set at 0.09~fb~(0.18~fb).

\begin{table}[tbp]
\caption{Results of the fits in the single-bin signal regions.  Left to right: The first column indicates the single-bin SRs under study.  The next two columns present observed and expected event yields in the single-bin regions, $N_{\mathrm{obs}}$ and $N_{\mathrm{exp}}$.  The latter are obtained after the control-region fit, and the errors include both the statistical and systematic uncertainties. The next two columns show the observed 95\% CL upper limits on the visible cross-section, $\langle\epsilon\sigma\rangle_{\mathrm{obs}}^{95}$, and on the number of signal events, $S_{\mathrm{obs}}^{95}$.  The next column shows the 95\% CL upper limit on the number of signal events, $S_{\mathrm{exp}}^{95}$, given the expected number (and $\pm 1\sigma$ deviations from the expectation) of background events. The last column indicates the discovery $p$-value, $p(s = 0)$.}
\renewcommand{\arraystretch}{1.2}
\label{tab:discovery}
\begin{tabular*}{\textwidth}{@{\extracolsep{\fill}}lcccccc}
\toprule
\toprule{\textrm{Single-Bin Region}} &$N_{\mathrm{obs}}$  &$N_{\mathrm{exp}}$ & $\langle\epsilon\sigma\rangle_{\textrm{obs}}^{95}$~[fb] & $S_{\textrm{obs}}^{95}$ & $S_{\textrm{exp}}^{95}$ & $p(s=0)$ \\
\midrule
\SRtwoj      &50 & 55.9 $\pm$ 3.7 & $0.09$ &  $13$ & $ { 18 }^{ +7 }_{ -5 }$ & 0.50 \\
\SRthreej    &44 & 39.8 $\pm$ 4.3 & $0.18$ &  $25$ & $ { 19 }^{ +9 }_{ -6 }$ & 0.19 \\
\bottomrule
\bottomrule
\end{tabular*}
\end{table}


In the absence of significant excesses in these bins, a fit configuration, referred to as the exclusion fit, is formed by extending the background-only fit to include all of the BDT score bins within \SRtwoj{} and \SRthreej.  Figure~\ref{fig:SR} shows the distributions of BDT scores in the signal regions after a fit using this configuration under the assumption that the only contributions are due to SM processes. The corresponding event yields in each SR bin are broken down by process in Tables~\ref{tab:SR2_yields} and \ref{tab:SR3_yields}. The extracted normalisation factors for the \Vjets backgrounds in this fit are found to agree within uncertainties with those from the background-only fit that only includes the CRs as constraining regions. Again, no significant excess of events is observed above the SM predictions.

\newcolumntype{H}{@{}>{\lrbox0}l<{\endlrbox}}

\begin{table}[tbp]
\caption{Observed event yields and fit results for \SRtwoj using the exclusion fit. The category `Other' contains rare backgrounds from diboson, triboson, and top-quark production processes. Uncertainties in the fitted background estimates combine statistical and systematic uncertainties.}
\begin{adjustbox}{max size={\textwidth}{\textheight}}
\begin{tabular}{l S[table-format=4.0(1)] S[table-format=4.0(1)] S[table-format=4.0(1)] S[table-format=4.1(1)] S[table-format=4.1(1)] S[table-format=4.1(1)] S[table-format=4.1(1)] S[table-format=4.1(1)]}
\toprule
\toprule
BDT Score & { [0.60, 0.64)} & { [0.64, 0.68)} & { [0.68, 0.72)} & { [0.72, 0.76)} & { [0.76, 0.80)} & { [0.80, 0.84)} & { [0.84, 0.88)} & { [0.88, 1.0]} \\
\midrule
Observed events          & 3712 & 2774 & 1946 & 1441 & 1054 & 729 & 326 & 50 \\
\midrule
Fitted SM events        & 3590 \pm 40 & 2806 \pm 32 & 1953 \pm 24 & 1490 \pm 20 & 1100 \pm 17 & 774 \pm 13 & 347 \pm 9 & 54.4 \pm 2.5 \\
\midrule
$Z$ + jets, QCD     & 2104 \pm 34 & 1648 \pm 28 & 1117 \pm 22 & 826 \pm 15 & 576 \pm 12 & 387 \pm 10 & 145 \pm 6 & 16.8 \pm 1.1 \\
$W$ + jets, QCD     & 993 \pm 30 & 691 \pm 23 & 422 \pm 18 & 292 \pm 11 & 190 \pm 10 & 108 \pm 7 & 36.6 \pm 3.4 & 3.7 \pm 0.7 \\
$Z$ + jets, EWK        & 322 \pm 8 & 307 \pm 6 & 273 \pm 5 & 249 \pm 5 & 224 \pm 6 & 192 \pm 6 & 117 \pm 5 & 25.4 \pm 1.9 \\
$W$ + jets, EWK         & 139 \pm 3 & 131 \pm 3 & 117 \pm 3 & 109 \pm 3 & 98 \pm 3 & 79 \pm 3 & 45 \pm 2 & 8.1 \pm 0.6 \\
Other            & 30 \pm 7 & 28 \pm 7 & 23 \pm 5 & 14.0 \pm 3.2 & 12.3 \pm 3.0 & 7.2 \pm 1.6 & 2.8 \pm 0.7 & 0.4 \pm 0.1 \\
Multijet         & {$0.5_{-0.5}^{+1.6}$}          & {$0.6_{-0.6}^{+1.9}$}          & {$0.6_{-0.6}^{+1.8}$}          & {$0.5_{-0.5}^{+1.4}$}          & {$0.3_{-0.3}^{+1.0}$}          & {$0.2_{-0.2}^{+0.5}$}          & {$<0.1$}          & {$<0.1$}          \\
\bottomrule
\bottomrule
\end{tabular}
\end{adjustbox}
\label{tab:SR2_yields}
\end{table}


\begin{table}[tbp]
\caption{Observed event yields and fit results for \SRthreej using the exclusion fit. The category `Other' contains rare backgrounds from diboson, triboson, and top-quark production processes. Uncertainties in the fitted background estimates combine statistical and systematic uncertainties.}
\begin{adjustbox}{max size={\textwidth}{\textheight}}
\begin{tabular}{l S[table-format=4.0(1)] S[table-format=4.0(1)] S[table-format=4.0(1)] S[table-format=4.1(1)] S[table-format=4.1(1)] S[table-format=4.1(1)] S[table-format=4.1(1)] S[table-format=4.1(1)]}
\toprule
\toprule
BDT Score & { [0.60, 0.64)} & { [0.64, 0.68)} & { [0.68, 0.72)} & { [0.72, 0.76)} & { [0.76, 0.80)} & { [0.80, 0.84)} & { [0.84, 0.88)} & { [0.88, 1.0]} \\
\midrule
Observed events         & 5504              & 4295              & 3068              & 2358              & 1611              & 936              & 398              & 44                    \\
\midrule
Fitted SM events      & 5530 \pm 60          & 4310 \pm 50          & 3120 \pm 40          & 2344 \pm 34          & 1570 \pm 25          & 904 \pm 18          & 358 \pm 10          & 36.6 \pm 2.6          \\
\midrule
$Z$ + jets, QCD & 3530 \pm 80          & 2740 \pm 60          & 1980 \pm 40          & 1481 \pm 33          & 974 \pm 24          & 564 \pm 18          & 202 \pm 10          & 21.3 \pm 2.1          \\
$W$ + jets, QCD & 1390 \pm 60          & 1050 \pm 40          & 708 \pm 29          & 503 \pm 21          & 310 \pm 17          & 138 \pm 8          & 55 \pm 5          & 3.6 \pm 0.7          \\
$Z$ + jets, EWK     & 249 \pm 5          & 226 \pm 4          & 204 \pm 5          & 177 \pm 4          & 154 \pm 4          & 121 \pm 4          & 68 \pm 3          & 6.7 \pm 0.8          \\
$W$ + jets, EWK     & 129 \pm 5          & 111 \pm 4          & 93 \pm 3          & 77 \pm 2          & 56 \pm 2          & 42 \pm 2          & 19 \pm 1          & 2.9 \pm 0.4          \\
Other           & 220 \pm 50          & 170 \pm 40          & 131 \pm 30          & 103 \pm 24          & 74 \pm 18          & 40 \pm 10          & 14 \pm 4          & 2.0 \pm 0.5          \\
Multijet        & {$10_{-10}^{+50}$}          & {$10_{-10}^{+31}$}          & {$6_{-6}^{+19}$}          & {$3_{-3}^{+10}$}          & {$1_{-1}^{+4}$}          & {$0.3_{-0.3}^{+1.1}$}          & {$<0.1$}          & {$0.1_{-0.1}^{+0.4}$}          \\
\bottomrule
\bottomrule
\end{tabular}
\end{adjustbox}
\label{tab:SR3_yields}
\end{table}


The CL$_\text{s}$ prescription in conjunction with the exclusion fit configuration is also used to perform hypothesis tests of specific SUSY scenarios to set limits on the class of simplified models used to optimise the search.  Exclusions at 95\% CL are presented in a two-dimensional plane with the horizontal axis given by the mass of the \chitz{}, and the vertical axis defined by the difference in mass between the \chitz{} and \chioz{}, and are shown in Figure~\ref{fig:limits}.
The presented mass limits are restricted to $\Delta m(\chitz/\chiopm,\chioz) \ge 0.2~\GeV$ to match the smallest available mass splittings in the simulated signal samples.
Since the analysis strategy does not depend on the reconstruction of the $\chitz/\chiopm$ decay products, the results are expected to remain valid for mass splittings below $0.2~\GeV$ as well, assuming that the potentially long lifetime of the $\chiopm$ does not degrade the signal acceptance.
The observed limit on \chitz{} masses for mass splittings below 1~\GeV is approximately \WinoMaxLimit, depending only slightly on the mass splitting.  This is greater than the expected limit of between 100 and 104~\GeV{} for the same $\Delta m(\chitz/\chiopm,\chioz)$ range due to the mild deficit of observed events in the high BDT score bins of \SRtwoj{}.  The \SRtwoj{} behavior is partially mitigated by the opposite trend in \SRthreej, where the high BDT score bins show an excess.  The use of separate normalisation factors for \Vjets backgrounds in the \twoj and \threej channels reduces the correlation between the two signal regions, allowing both channels to influence the limits through their sensitivity to the benchmark signal models.

\begin{figure}[tbp]
\centering
\subfloat[]{
\includegraphics[width=0.48\textwidth]{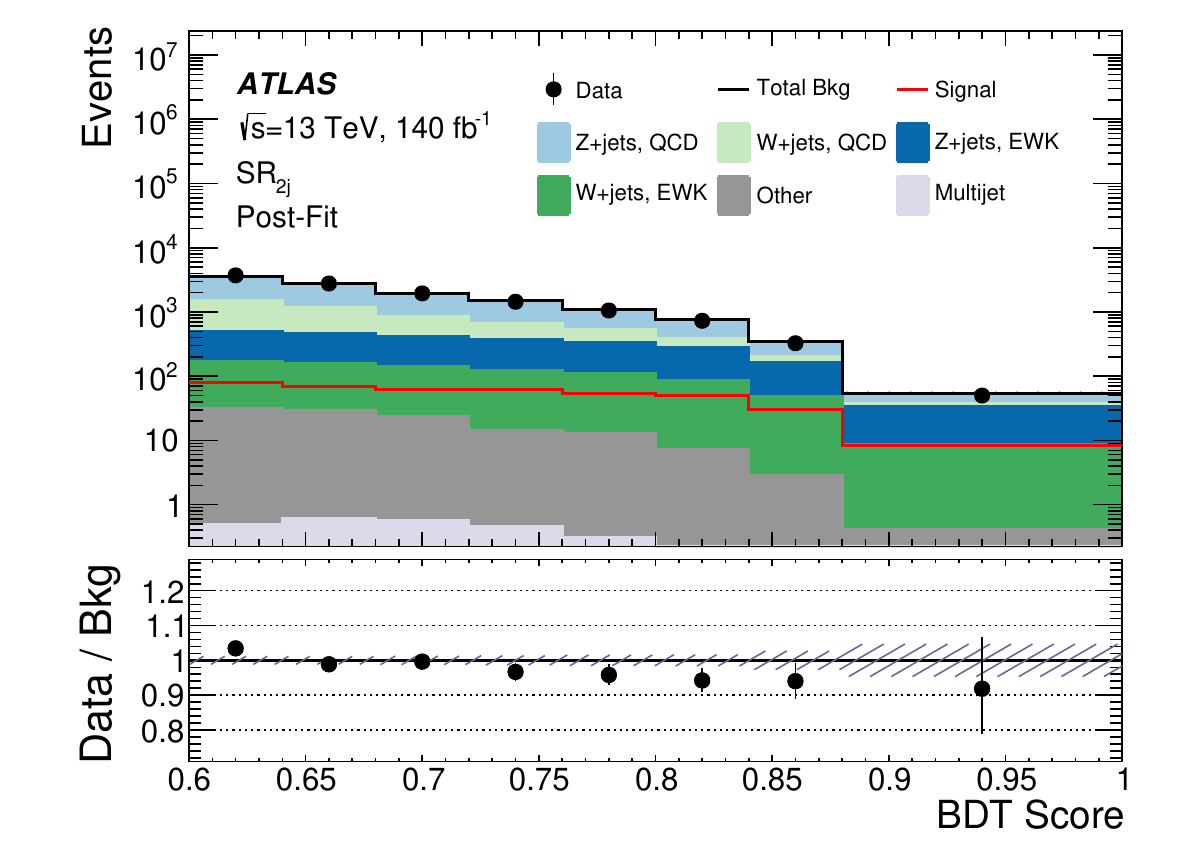}
}
\subfloat[]{
\includegraphics[width=0.48\textwidth]{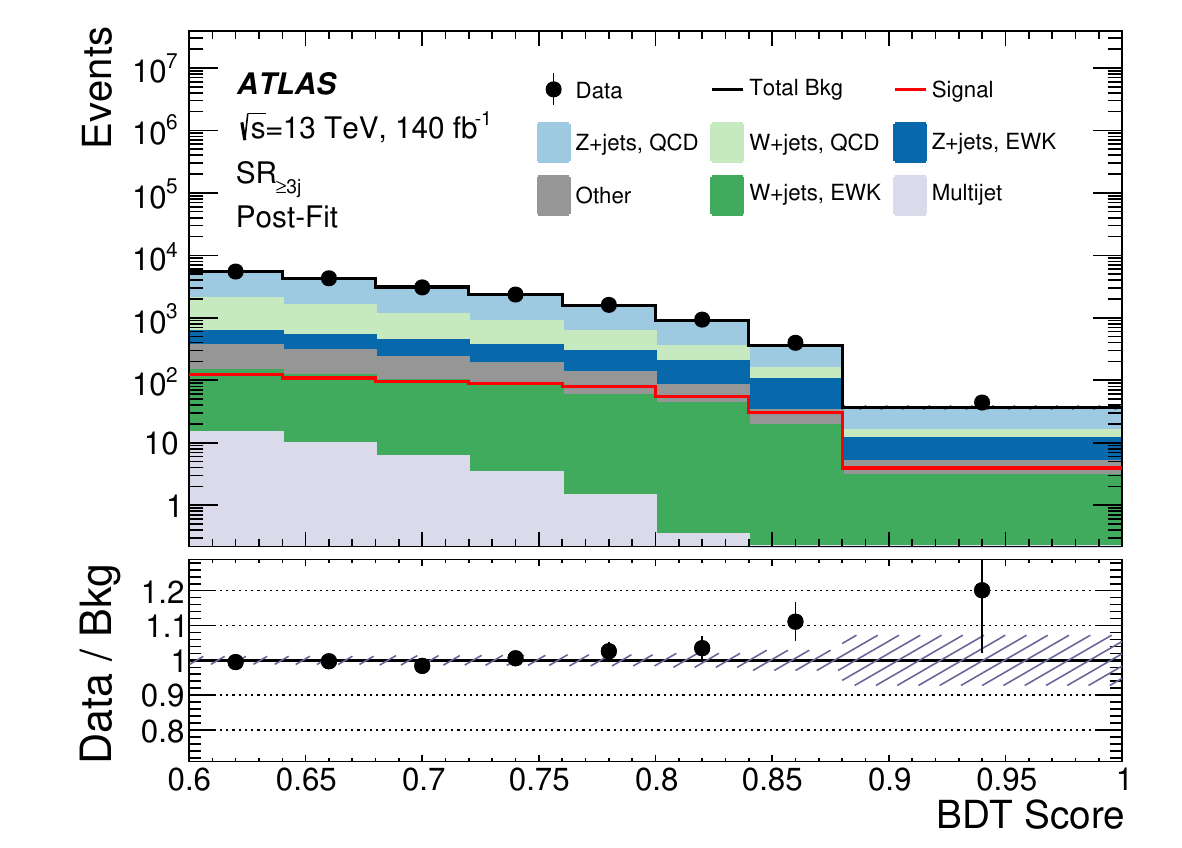}
}
\caption{Observed and predicted background distributions of the BDT score in (a) \SRtwoj{} and (b) \SRthreej{} after the exclusion fit. The nominal, pre-fit prediction of an example benchmark signal with $(m(\chitz/\chiopm), \mchioz) = (100, 99)~\GeV$ is shown in red. The `Other' category contains rare backgrounds from diboson, triboson and top-quark production processes. The hatched band represents the post-fit experimental, theoretical, and statistical uncertainties in the total background. The bottom panel of each plot shows the ratio between the data and the post-fit background prediction.}
\label{fig:SR}
\end{figure}

\begin{figure}[tbp]
\centering
\includegraphics[width=0.8\columnwidth]{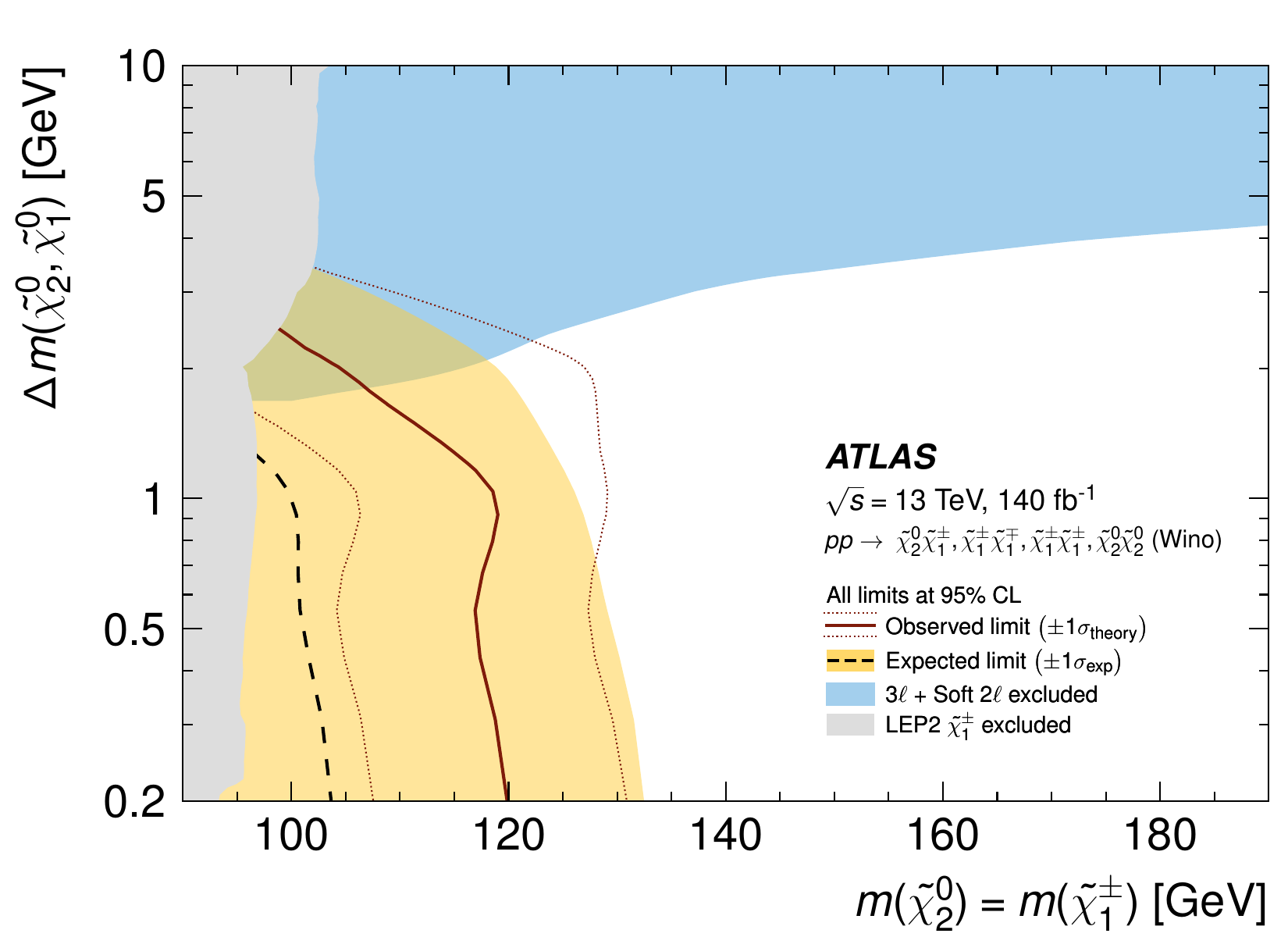}
\caption{Expected (dashed black line) and observed (solid red line) 95\% CL exclusion limits on the compressed SUSY simplified model with a bino-like LSP and wino-like NLSPs being considered. These are shown with $\pm1\sigma_\text{exp}$ (yellow band) from experimental systematic and statistical uncertainties, and with $\pm1\sigma^{\text{SUSY}}_{\text{theory}}$ (red dotted lines) from signal cross-section uncertainties, respectively. The limits set by the ATLAS searches using the soft lepton~\cite{SUSY-2018-16,SUSY-2019-09} signature is illustrated by the blue region while the limit imposed by the LEP experiments~\cite{LEPlimits} is shown in grey.
}
\label{fig:limits}
\end{figure}

\FloatBarrier

\section{Conclusion}

A search for beyond the SM physics is performed using events consistent with the production of invisible particles
via VBF processes using $\sqrt{s}=13~\TeV$ $pp$ collision data corresponding to an integrated luminosity of 140~\ifb
recorded by the ATLAS detector CERN's LHC\@.
The events are required to contain at least two jets with a large separation in pseudorapidity and dijet invariant mass, large
missing transverse momentum, and no reconstructed leptons.
The events are further classified into two categories that require exactly two or at least three reconstructed jets.
Discrimination between SM backgrounds and signal is enhanced by the use of a BDT that is trained on a simplified SUSY
model in which a pair of mass-degenerate, wino-like charginos and neutralinos is produced in association with jets
before decaying into a close-in-mass ($\Delta m(\chitz/\chiopm,\chioz) \lesssim 1~\GeV$) bino-like neutralino LSP and soft SM particles.
This strategy provides unique sensitivity to compressed SUSY scenarios where existing searches relying on soft-lepton signatures become insensitive due to limitations in the reconstruction thresholds for leptons.
Unlike those searches, the final state targeted here also allows this search to remain agnostic to the branching ratios of the $\chitz/\chiopm$ decays.
The data are found to be consistent with the SM predictions and upper limits are set at 95\% CL on the visible cross-section of generic, non-SM
processes in the targeted phase space and on the chargino and neutralino masses in a simplified SUSY model featuring wino-like $\chitz/\chiopm$ states that are nearly degenerate in mass with the bino-like $\chioz$.
Including the effects of interference that impact the total signal cross-section and kinematics, a lower limit of \WinoMaxLimit is set on
the wino-like $\chitz$ and $\chiopm$ masses when they are within $1~\GeV$ of the bino-like $\chioz$ mass, surpassing previously constraints from the LEP experiment by up to approximately 25~\GeV.

\section*{Acknowledgements}
%

%
%

%
%

We thank CERN for the very successful operation of the LHC and its injectors, as well as the support staff at
CERN and at our institutions worldwide without whom ATLAS could not be operated efficiently.

The crucial computing support from all WLCG partners is acknowledged gratefully, in particular from CERN, the ATLAS Tier-1 facilities at TRIUMF/SFU (Canada), NDGF (Denmark, Norway, Sweden), CC-IN2P3 (France), KIT/GridKA (Germany), INFN-CNAF (Italy), NL-T1 (Netherlands), PIC (Spain), RAL (UK) and BNL (USA), the Tier-2 facilities worldwide and large non-WLCG resource providers. Major contributors of computing resources are listed in Ref.~\cite{ATL-SOFT-PUB-2023-001}.

We gratefully acknowledge the support of ANPCyT, Argentina; YerPhI, Armenia; ARC, Australia; BMWFW and FWF, Austria; ANAS, Azerbaijan; CNPq and FAPESP, Brazil; NSERC, NRC and CFI, Canada; CERN; ANID, Chile; CAS, MOST and NSFC, China; Minciencias, Colombia; MEYS CR, Czech Republic; DNRF and DNSRC, Denmark; IN2P3-CNRS and CEA-DRF/IRFU, France; SRNSFG, Georgia; BMBF, HGF and MPG, Germany; GSRI, Greece; RGC and Hong Kong SAR, China; ISF and Benoziyo Center, Israel; INFN, Italy; MEXT and JSPS, Japan; CNRST, Morocco; NWO, Netherlands; RCN, Norway; MNiSW, Poland; FCT, Portugal; MNE/IFA, Romania; MSTDI, Serbia; MSSR, Slovakia; ARIS and MVZI, Slovenia; DSI/NRF, South Africa; MICIU/AEI, Spain; SRC and Wallenberg Foundation, Sweden; SERI, SNSF and Cantons of Bern and Geneva, Switzerland; NSTC, Taipei; TENMAK, T\"urkiye; STFC/UKRI, United Kingdom; DOE and NSF, United States of America.

Individual groups and members have received support from BCKDF, CANARIE, CRC and DRAC, Canada; CERN-CZ, FORTE and PRIMUS, Czech Republic; COST, ERC, ERDF, Horizon 2020, ICSC-NextGenerationEU and Marie Sk{\l}odowska-Curie Actions, European Union; Investissements d'Avenir Labex, Investissements d'Avenir Idex and ANR, France; DFG and AvH Foundation, Germany; Herakleitos, Thales and Aristeia programmes co-financed by EU-ESF and the Greek NSRF, Greece; BSF-NSF and MINERVA, Israel; NCN and NAWA, Poland; La Caixa Banking Foundation, CERCA Programme Generalitat de Catalunya and PROMETEO and GenT Programmes Generalitat Valenciana, Spain; G\"{o}ran Gustafssons Stiftelse, Sweden; The Royal Society and Leverhulme Trust, United Kingdom.

In addition, individual members wish to acknowledge support from Armenia: Yerevan Physics Institute (FAPERJ); CERN: European Organization for Nuclear Research (CERN PJAS); Chile: Agencia Nacional de Investigaci\'on y Desarrollo (FONDECYT 1230812, FONDECYT 1230987, FONDECYT 1240864); China: Chinese Ministry of Science and Technology (MOST-2023YFA1605700), National Natural Science Foundation of China (NSFC - 12175119, NSFC 12275265, NSFC-12075060); Czech Republic: Czech Science Foundation (GACR - 24-11373S), Ministry of Education Youth and Sports (FORTE CZ.02.01.01/00/22\_008/0004632), PRIMUS Research Programme (PRIMUS/21/SCI/017); EU: H2020 European Research Council (ERC - 101002463); European Union: European Research Council (ERC - 948254, ERC 101089007), European Union, Future Artificial Intelligence Research (FAIR-NextGenerationEU PE00000013), Italian Center for High Performance Computing, Big Data and Quantum Computing (ICSC, NextGenerationEU); France: Agence Nationale de la Recherche (ANR-20-CE31-0013, ANR-21-CE31-0013, ANR-21-CE31-0022, ANR-22-EDIR-0002); Germany: Baden-Württemberg Stiftung (BW Stiftung-Postdoc Eliteprogramme), Deutsche Forschungsgemeinschaft (DFG - 469666862, DFG - CR 312/5-2); Italy: Istituto Nazionale di Fisica Nucleare (ICSC, NextGenerationEU), Ministero dell'Università e della Ricerca (PRIN - 20223N7F8K - PNRR M4.C2.1.1); Japan: Japan Society for the Promotion of Science (JSPS KAKENHI JP22H01227, JSPS KAKENHI JP22H04944, JSPS KAKENHI JP22KK0227, JSPS KAKENHI JP23KK0245); Netherlands: Netherlands Organisation for Scientific Research (NWO Veni 2020 - VI.Veni.202.179); Norway: Research Council of Norway (RCN-314472); Poland: Ministry of Science and Higher Education (IDUB AGH, POB8, D4 no 9722), Polish National Agency for Academic Exchange (PPN/PPO/2020/1/00002/U/00001), Polish National Science Centre (NCN 2021/42/E/ST2/00350, NCN OPUS 2023/51/B/ST2/02507, NCN OPUS nr 2022/47/B/ST2/03059, NCN UMO-2019/34/E/ST2/00393, NCN \& H2020 MSCA 945339, UMO-2020/37/B/ST2/01043, UMO-2021/40/C/ST2/00187, UMO-2022/47/O/ST2/00148, UMO-2023/49/B/ST2/04085, UMO-2023/51/B/ST2/00920); Slovenia: Slovenian Research Agency (ARIS grant J1-3010); Spain: Generalitat Valenciana (Artemisa, FEDER, IDIFEDER/2018/048), Ministry of Science and Innovation (MCIN \& NextGenEU PCI2022-135018-2, MICIN \& FEDER PID2021-125273NB, RYC2019-028510-I, RYC2020-030254-I, RYC2021-031273-I, RYC2022-038164-I); Sweden: Carl Trygger Foundation (Carl Trygger Foundation CTS 22:2312), Swedish Research Council (Swedish Research Council 2023-04654, VR 2018-00482, VR 2022-03845, VR 2022-04683, VR 2023-03403, VR grant 2021-03651), Knut and Alice Wallenberg Foundation (KAW 2018.0458, KAW 2019.0447, KAW 2022.0358); Switzerland: Swiss National Science Foundation (SNSF - PCEFP2\_194658); United Kingdom: Leverhulme Trust (Leverhulme Trust RPG-2020-004), Royal Society (NIF-R1-231091); United States of America: U.S. Department of Energy (ECA DE-AC02-76SF00515), Neubauer Family Foundation.

%
%


%
\clearpage
\printbibliography
\clearpage
 
\begin{flushleft}
\hypersetup{urlcolor=black}
{\Large The ATLAS Collaboration}

\bigskip

\AtlasOrcid[0000-0002-6665-4934]{G.~Aad}$^\textrm{\scriptsize 105}$,
\AtlasOrcid[0000-0001-7616-1554]{E.~Aakvaag}$^\textrm{\scriptsize 17}$,
\AtlasOrcid[0000-0002-5888-2734]{B.~Abbott}$^\textrm{\scriptsize 124}$,
\AtlasOrcid[0000-0002-0287-5869]{S.~Abdelhameed}$^\textrm{\scriptsize 120a}$,
\AtlasOrcid[0000-0002-1002-1652]{K.~Abeling}$^\textrm{\scriptsize 57}$,
\AtlasOrcid[0000-0001-5763-2760]{N.J.~Abicht}$^\textrm{\scriptsize 51}$,
\AtlasOrcid[0000-0002-8496-9294]{S.H.~Abidi}$^\textrm{\scriptsize 30}$,
\AtlasOrcid[0009-0003-6578-220X]{M.~Aboelela}$^\textrm{\scriptsize 46}$,
\AtlasOrcid[0000-0002-9987-2292]{A.~Aboulhorma}$^\textrm{\scriptsize 36e}$,
\AtlasOrcid[0000-0001-5329-6640]{H.~Abramowicz}$^\textrm{\scriptsize 156}$
% ...
\bigskip
\\

% ...
$^{17}$Department for Physics and Technology, University of Bergen, Bergen; Norway.\\
% ...
$^{30}$Physics Department, Brookhaven National Laboratory, Upton NY; United States of America.\\
% ...
$^{36}$$^{(a)}$Facult\'e des Sciences Ain Chock, Universit\'e Hassan II de Casablanca;$^{(b)}$Facult\'{e} des Sciences, Universit\'{e} Ibn-Tofail, K\'{e}nitra;$^{(c)}$Facult\'e des Sciences Semlalia, Universit\'e Cadi Ayyad, LPHEA-Marrakech;$^{(d)}$LPMR, Facult\'e des Sciences, Universit\'e Mohamed Premier, Oujda;$^{(e)}$Facult\'e des sciences, Universit\'e Mohammed V, Rabat;$^{(f)}$Institute of Applied Physics, Mohammed VI Polytechnic University, Ben Guerir; Morocco.\\
% ...
$^{46}$Physics Department, Southern Methodist University, Dallas TX; United States of America.\\
% ...
$^{51}$Fakult\"{a}t Physik , Technische Universit{\"a}t Dortmund, Dortmund; Germany.\\
% ...
$^{57}$II. Physikalisches Institut, Georg-August-Universit\"{a}t G\"ottingen, G\"ottingen; Germany.\\
% ...
$^{105}$CPPM, Aix-Marseille Universit\'e, CNRS/IN2P3, Marseille; France.\\
% ...
$^{120}$$^{(a)}$New York University Abu Dhabi, Abu Dhabi;$^{(b)}$United Arab Emirates University, Al Ain; United Arab Emirates.\\
% ...
$^{124}$Homer L. Dodge Department of Physics and Astronomy, University of Oklahoma, Norman OK; United States of America.\\
% ...
$^{156}$Raymond and Beverly Sackler School of Physics and Astronomy, Tel Aviv University, Tel Aviv; Israel.\\
% ...
$^{a}$ Also Affiliated with an institute covered by a cooperation agreement with CERN.\\
$^{b}$ Also at An-Najah National University, Nablus; Palestine.\\
$^{c}$ Also at Borough of Manhattan Community College, City University of New York, New York NY; United States of America.\\
$^{d}$ Also at Center for High Energy Physics, Peking University; China.\\
$^{e}$ Also at Center for Interdisciplinary Research and Innovation (CIRI-AUTH), Thessaloniki; Greece.\\
$^{f}$ Also at CERN, Geneva; Switzerland.\\
% ...

\end{flushleft}


%

%
%

%

%
%
%
%

\end{document}